\documentclass[numbers,preprint,11pt]{elsarticle}
\journal{Computer Physics Communications}

\usepackage{amsmath,amssymb}
\usepackage[english]{babel}
\usepackage[T1]{fontenc}

\usepackage[colorlinks,linkcolor=blue,citecolor=blue,urlcolor=blue]{hyperref}
\hypersetup{pdftitle={A new hybrid code (CHIEF) implementing the inertial electron fluid equation without approximation},pdfauthor={P. A. Mu\~noz}}

\biboptions{numbers}

\usepackage{textcomp}
\begin{document}

\begin{frontmatter}
\title{A new hybrid code (CHIEF) implementing the inertial electron fluid equation without approximation
	\footnote{
	\textcopyright 2018. This manuscript version is made available under the CC-BY-NC-ND 4.0 license \href{http://creativecommons.org/licenses/by-nc-nd/4.0/}{http://creativecommons.org/licenses/by-nc-nd/4.0/}
	}
}

\author[mps,mppc]{P. A. Mu\~noz}
\ead{munozp@mps.mpg.de}
\author[mps,mppc]{N. Jain}
\author[potch]{P. Kilian}
\author[mps,mppc]{J. B\"uchner}
\address[mps]{Max-Planck-Institut f\"ur Sonnensystemforschung, D-37077 G\"ottingen, Germany}
\address[mppc]{Max-Planck/Princeton Center for Plasma Physics}
\address[potch]{Centre for Space Research, North-West University, 2520 Potchefstroom, South Africa}
\date{\today}

\begin{abstract}
	We present a new hybrid algorithm implemented in the code CHIEF (\underline{C}ode \underline{H}ybrid with \underline{I}nertial \underline{E}lectron \underline{F}luid) for simulations of electron-ion plasmas.
	The algorithm treats the ions kinetically, modeled by the Particle-in-Cell (PiC) method,
	and electrons as an inertial fluid, modeled by electron fluid equations without any of the approximations
	used in most of the other hybrid codes with an inertial electron fluid.
	This kind of code is appropriate to model a large variety of quasineutral plasma phenomena
	where the electron inertia and/or ion kinetic effects are relevant.
	We present here the governing equations of the model, how these are discretized and implemented numerically, as well as six test problems to validate our numerical approach.
	Our chosen test problems, where the electron inertia and ion kinetic effects play the essential role,
	are:
	0) Excitation of parallel eigenmodes to check numerical convergence and stability,
	1) parallel (to a background magnetic field) propagating electromagnetic waves,
	2) perpendicular propagating electrostatic waves (ion Bernstein modes),
	3) ion beam right-hand instability (resonant and non-resonant),
	4) ion Landau damping,
	5) ion firehose instability,
	and
	6) 2D oblique ion firehose instability.
	Our results reproduce successfully the predictions of linear and non-linear theory for all these problems, validating our code.
	All properties of this hybrid code make it  ideal
	to study multi-scale phenomena between electron and ion scales such as
	collisionless shocks, magnetic reconnection and kinetic plasma turbulence in the dissipation range above the electron scales.

\end{abstract}


\end{frontmatter}

\section{\label{sec:intro}Introduction}

Fully kinetic simulations, either via Lagrangian Particle-in-Cell (PiC)~\cite{Birdsall1991,Hockney1988,Buchner2003} or Eulerian Vlasov~\cite{Mangeney2002,Filbet2003,Umeda2012} methods, have become the standard tools to study kinetic plasma phenomena. They are essential to model self-consistently the dissipation at small scales and  the multiscale processes in a wide variety of scenarios, often related with turbulence. But these simulations codes tend to be computationally very expensive, which hinders their applicability to many practical problems.

On the other hand, fluid models are used to model phenomena either on large scales and long timescales, like Magnetohydrodynamics (MHD) codes, or even on small space and timescales under appropriate conditions, like  electron-MHD (EMHD) codes~\cite{Drake1994,Jain2003}. Even though they are computationally more convenient than kinetic models, they cannot model self-consistently dissipation, particle acceleration, resonant processes or phenomena related with deviations from the thermal equilibrium or quasineutrality.

Hybrid codes bridge the gap between those different approaches and spatial/time scales. A large number of problems require kinetic effects of ions and fluid descriptions of electrons. This is especially true at frequencies lower than the ion cyclotron $\Omega_{ci}$ or spatial scales much larger than the ion skin depth $d_i$, which are computationally expensive to simulate with fully-kinetic codes. Problems of this class are related to planetary magnetospheres and their associated collisionless shocks, collisionless magnetic reconnection, the acceleration of heavy ions by shocks and reconnection, instabilities driven by ion temperature anisotropy and turbulence in the kinetic dissipation range in the solar wind, among many other ones. In many of these scenarios, the gradient scales of different physical quantities  can be very small, which requires to consider an electron fluid with finite mass.

In order to study these problems, we present a hybrid code using kinetic ions modeled via PiC methods, and a massive electron fluid using a modified EMHD model. We call it
CHIEF: Code Hybrid with Inertial Electron Fluid.
By not considering electron kinetic effects and going to the radiation-free limit, we can bypass the stringent stability conditions of a fully kinetic approach, making it computationally cheaper. Our code follows the standard quasineutral approach used in kinetic hybrid codes, i.e., the electron and ion densities are always equal, eliminating thus the radiation term in the Maxwell's equation (Amp\`ere's) associated with the displacement current, light wave propagation (and the corresponding CFL condition of standard fully-kinetic codes) and other phenomena involving charge separation such as Langmuir waves.
On the other hand, and different from other hybrid codes with inertial electrons (see, e.g., the review Ref.~\cite{Winske2003} or the textbook Ref.~\cite{Lipatov2002} and references therein), we consider without approximations all the electron inertial terms in the generalized Ohm's law. Note that the inclusion of electron inertia reduces the numerical instability (at short wavelengths) known for hybrid codes with massless electrons.
This instability is due to the increase of the phase speed of whistler waves without bound, since $\omega\propto k^2$ in that limit, where $\omega$ is the frequency and $k$ the wave number. All these features make our code ideal to deal with fast evolving phenomena at electron scales, which can break the frozen-in condition of ideal MHD allowing processes like magnetic reconnection.

Note that our main approximation is the choice of constant temperature as  equation of state for the electrons.
Other equations of state such as the polytropic one, where the electron pressure has an explicit dependence on other variables such as the density, can be straightforwardly implemented. On the other hand, numerical solutions to equations for the evolution of the electron pressure require additional numerical considerations, like adding a numerical viscosity, see Ref.~\cite{Shay1998}. That is especially true, in particular, when non-gyrotropic contributions are considered in the electron pressure tensor, since it makes necessary to include a heuristic isotropization term to take into account pitch angle scattering due to electron temperature anisotropy driven instabilities (see Ref.~\cite{Hesse1995}).

In order to make the advantages of our code clear, let us briefly discuss the different approaches that previous hybrid codes have taken to implement the relevant electron fluid equations and their coupling with the Maxwell's equations. Even since the early days of the development of (PiC-)hybrid codes, electron inertia has been considered to deal with fast phenomena close to the electron cyclotron frequency, as demonstrated by an early 1D electrostatic code~\cite{Forslund1971}. A 2.5D electromagnetic hybrid code was developed by Ref.~\cite{Hewett1978}, allowing some electron mass effects. Those codes~\cite{Forslund1971,Hewett1978} implemented the quasineutral approximation by using the radiation-free (Darwin) limit of the Maxwell's equations. Correspondingly, they used a Helmholtz decomposition by solving separately the longitudinal and transverse parts of the electromagnetic potentials and current density. Longitudinal/transverse refers to the curl-free/divergence-free  part of the corresponding fields, and not to physical directions in space. The electron inertia was considered in the equation for the transverse electron current density. An evolution equation for the electron temperature including terms up to the heat flux was considered. This method involved solving several elliptic equations for the fields (with the form of generalized Poisson's equations). Hybrid codes based on potentials, however, have been rarely used since then, except 1D cases which use the electrostatic potential. One example is the 1D code developed in Ref.~\cite{Eremin2016}, which modeled a highly collisional plasma typical of pressure discharges with a complex chemistry. It consists of fully-kinetic electrons (using Monte-Carlo methods to take into account particle collisions) and several species of heavy ions modeled with a fluid approach.

Much later, Ref.~\cite{Swift1996} developed a (2.5D) hybrid code algorithm to simulate the entire Earth's magnetosphere. To do so, it included a cold ion fluid component to take into account the ionospheric plasma, as well as curvilinear coordinates. In this approach, the electric field is determined from the electron momentum equation with electron inertia, while the electron velocity from the Amp\`ere's law and the magnetic field from the Faraday's law. No electron pressure term was considered in the electron momentum equation. The electron inertia term was only considered as a correction in the form of an electron polarization drift in the (implicit) equation for the magnetic field, while neglected in all the other equations. This has the advantage, however, of making the entire code explicit. This code was applied to the substorm onset and other magnetospheric scenarios in Ref.~\cite{Swift2001}.

After that, the standard approach in inertial hybrid codes has been to solve equations for generalized electromagnetic fields (see Sec.~5.7 in Ref.~\cite{Lipatov2002}, Refs.~\cite{Shay1998,Kuznetsova1998}).
In this approach, the electromagnetic fields are obtained from a generalized magnetic $\hat{\vec{B}}$ and electric  $\hat{\vec{E}}$ fields which satisfy a Faraday's-like equation.
Those 2.5D hybrid codes have been used to study collisionless magnetic reconnection~\cite{Shay1999,Kuznetsova2000,Kuznetsova2001}.
The equations for the generalized fields  $\hat{\vec{B}}$ and  $\hat{\vec{E}}$ were solved by using a predictor-corrector scheme (which requires a staggered grid) or a trapezoidal leapfrog algorithm.
As discussed in detail later, the electromagnetic fields are obtained from the generalized fields under different approximations. Spatio-temporal density variations at electron scales and the time derivative of the ion current were neglected in the expressions for $\hat{\vec{B}}$ and $\hat{\vec{E}}$, respectively~\cite{Kuznetsova1998}.  Although the contribution of electron inertia via the convective electron acceleration term was considered in calculating the electric field $\hat{\vec{E}}$, the time-derivative of the electron inertia was neglected~\cite{Kuznetsova1998}. Some authors neglected even the convective electron acceleration term~\cite{Shay1998}.
On the other hand, Ref.~\cite{Shay1998} used an evolution equation for a scalar electron pressure, while  Ref.~\cite{Kuznetsova1998} included the full electron pressure tensor to take into account the non-gyrotropic effects that were shown to have an important role in balancing the reconnection electric field under some conditions. None of the effects related with an equation for the evolution of the electron pressure is considered in our code, yet.

A slightly different method to obtain the electric field from $\hat{\vec{E}}$ consists in writing implicitly the equation for the generalized electric field (instead of the magnetic field), combining the Ohm's law with the other Maxwell's equations, resulting in an elliptic equation for  $\vec{E}$ (see Sec.~5.2.4 in Ref.~\cite{Lipatov2002}). They pointed out that one of the main advantages of this approach is the possibility to simulate regions with very small density, something that the previous methods cannot handle easily due to the division by this quantity in several of the Ohm's law terms. This kind of hybrid codes was applied mainly for simulations of collisionless shocks.

In hybrid codes, ions can be modeled not only via PiC methods, but also via Eulerian Vlasov methods. Ref.~\cite{Valentini2007} developed a (2.5D) hybrid Vlasov code incorporating electron inertia. This code couples the ion equation of motion represented by the Vlasov equation, solved by standard splitting methods, with the Maxwell's equations via the current advance method (CAM)~\cite{Matthews1994}.
The CAM method uses the ion momentum equation (obtained by calculating the first-order momenta of the Vlasov equation) including a tensor pressure, in order to advance the ion current and thus  to calculate the electric field.
The corresponding generalized Ohm's law is a Helmholtz equation for the generalized electric field written in terms of the ion fluid velocity and the ion tensor pressure. It does not involve time derivatives, because of the quasineutral approximation and Faraday's law. The solution of this equation is simplified by assuming negligible density perturbations in the expression for the electric field. The equation of state for the electrons is assumed to be isothermal. The code by Ref.~\cite{Valentini2007} has been applied to several problems related to solar wind turbulence  (see, e.g., Refs.~\cite{Valentini2008,Valentini2010,Valentini2011}).

More recently, Ref.~\cite{Cheng2013} developed a (3D) hybrid code with electron inertia and kinetic ions modeled via the $\delta f$ method, by means of a second-order accurate  semi-implicit algorithm using a current closure scheme. The $\delta f$ method consists in evolving only the variations of the ion distribution function, which are assumed to be small compared to a given background, and thus to reduce the numerical noise predominant in conventional full-$f$ methods. The electron inertial effects are included in the generalized Ohm's law for the electric field,  and are calculated by using the Amp\`ere's law and the ion momentum equation. This approach was first proposed by Ref.~\cite{Jones2003} for the 1D gyrofluid electron equations, being later applied to study the propagation of dispersive Alfv\'en waves in the Earth's magnetosphere-ionosphere region and the associated mechanisms of electron acceleration~\cite{Su2004,Su2007}, and the Alfv\'en dynamics in an Io-Jupiter flux tube~\cite{Su2006}.
The electric field is advanced first by decomposing the Ohm's law into a perturbed and an equilibrium part. The fields are solved with finite differences in the inhomogeneous direction and with spectral methods in the homogeneous direction. The equilibrium part of the Ohm's law is cast in the form of a linear equation and solved by matrix inversion methods, while the perturbed part is solved iteratively. After that, the magnetic field is obtained explicitly by using Faraday's law and finally, the electromagnetic fields can be used to advance the (perturbed) particle distribution. The closure for the electron pressure was chosen to be an isothermal equation of state. This code was benchmarked against the (2D) resistive tearing instability, among other (1D) problems.

Finally, one of the more recent hybrid codes with electron inertia was published in Ref.~\cite{Amano2014}. The special feature of this 1D code is that can handle low density and vacuum regions without becoming unstable. In order to do that, a correction term is introduced to the electric instead of the magnetic field, in addition to a variable (ion-to-electron) mass ratio which reduces the phase speed of whistler waves to satisfy  the CFL stability condition everywhere. This is done not only because of the divisions by density in the equation for the electric field, but also because the Alfv\'en speed is inversely proportional to (the square root of) the density, implying a more stringent condition in the choice of the time step and the relevant CFL condition (associated to the electron Alfv\'en speed). The approach by Ref.~\cite{Amano2014} consisted in not using the generalized electromagnetic fields as in Refs.~\cite{Shay1998,Kuznetsova1998,Lipatov2002}, but instead solving a modified  form of the Ohm's law for the electric field (without divisions by density) and then obtaining the magnetic field through the Faraday's law. The first equation is solved by matrix methods while the second one by a direct iterative algorithm. The electron inertia effects are considered in both equations, however, with some approximations. In the vacuum limit, and following the idea of a previous inertia-less hybrid code~\cite{Harned1982}, the equation for the electric field becomes the Laplace's equation, being easily solvable without numerical singularities as in other approaches due to the division by density. The electron pressure is considered to follow an adiabatic equation of state, with a corresponding evolution equation in case of a finite resistivity (constant otherwise). Finally, the variable mass ratio procedure is done in practice by varying locally and temporally the electron mass, adjusting it to always satisfy the CFL condition based on the electron Alfv\'en speed for given timestep and grid cell sizes. This works well as long as the scales of interest are not too close to the electron inertial scales.

The rest of this paper is organized as follows. In Sec.~\ref{sec:model} we present a detailed description of the simulation model used by our code. In Sec~\ref{sec:numerical_implementation} we describe the numerical implementation of our simulation model. In Sec.~\ref{sec:tests} we show  a detailed comparison of simulations results of one numerical and six physical test problems with the predictions of linear (and non-linear) theory: 0) excitation of parallel eigenmodes
(to check numerical convergence and stability),
1) parallel (to a background magnetic field) propagating electromagnetic waves, 2) perpendicular propagating electrostatic waves (ion Bernstein modes), 3) ion beam right-hand instability (resonant and non-resonant), 4) ion Landau damping,
5) ion firehose instability
and
6) 2D oblique ion firehose instability.
These test problems cover different aspects of ion kinetic and electron inertial effects, essential to understand phenomena in a wide variety of laboratory, space and astrophysical plasmas. All the results match within the expected error range. We summarize our results in Sec.~\ref{sec:conclusion}.

\section{Simulation model\label{sec:model}}
We treat ions
as Lagrangian macro-particles modeled via the PiC method and electrons as a (Eulerian) fluid with finite mass and temperature.  Each ion macro-particle (index ``$p$'') represents a (large) number $N_i$ of physical ions (index ``$i$'') according to
$f_i^p(\vec{x}_i^{\;p},\vec{v}_i^{\;p},t)=N_i^p\,S(\vec{x}_i-\vec{x}_i^{\;p}(t))\delta(\vec{v}_i-\vec{v}_i^{\;p}(t))$, where
$\vec{x}_i^{\;p}$, $\vec{v}_i^{\;p}$  are the macro-ions' position, velocity and $S(\vec{x}_i-\vec{x}_i^{\;p}(t))$ is called the shape function (see details in Sec.~\ref{sec:numerical_implementation}). The  (physical)  ion distribution function is obtained as the summation over all ion macro-particles
$f_i(\vec{x}_i,\vec{v}_i,t)=\sum_p f_i^p(\vec{x}_i^{\;p},\vec{v}_i^{\;p},t) =\sum_p N_i^p\,S(\vec{x}_i-\vec{x}_i^{\;p}(t))\delta(\vec{v}_i-\vec{v}_i^{\;p}(t))$.
In the collisionless plasmas to be considered, this macro-ion distribution function $f_i^p$ evolves according to a Vlasov-like equation governing the full ion distribution function $f_i$ (see, e.g., Ref.~\cite{Lapenta2012}):
\begin{align}\label{eq:vlasov}
	\frac{\partial f_i^p}{\partial t} + \vec{v}_i\cdot \frac{\partial f_i^p}{\partial \vec{x}_i} + \frac{Ze}{m_i}\left(\vec{E} + \vec{v}_i\times \vec{B}\right)\cdot\frac{\partial f_i^p}{\partial \vec{v}_i}=0.
\end{align}
Here, $Ze$ and $m_i$ are the charge (with $e$ the fundamental charge) and mass of a physical ion, respectively. By taking the first order momenta  in $\vec{x}_i$ and $\vec{v}_i$ of Eq.~\eqref{eq:vlasov}  (i.e., multiplying and integrating the Vlasov equation by $\iint  \vec{x}_i\, {\rm (Eq.\,(1))}\,d\vec{x}_i\,d\vec{v}_i$ and $\iint \vec{v}_i {\rm (Eq.\,(1))} \,d\vec{x}_i\,d\vec{v}_i$, respectively. See details in, e.g., Ref.~\cite{Lapenta2012}), one obtains the equations of motion of the macro-ions (setting $Z=1$, i.e., singly charged ions, for simplicity):
\begin{align}
	\frac{d\vec{x}_i^{\;p}}{dt}    & =\vec{v}_i^{\;p}, \label{eq:xp}                                \\
	m_i\frac{d\vec{v}_i^{\;p}}{dt} & =e(\vec{E}_i^{\;p} + \vec{v}_i^{\;p}\times \vec{B}_i^{\;p}). \label{eq:vp}
\end{align}
where $\vec{x}_i^{\;p}$ and $\vec{v}_i^{\;p}$ are the first order momenta in $\vec{x}_i$ and $\vec{v}_i$ of $f_i^p$, respectively.
These equations resemble Newton's equation of motion for the macro-particles,
with the difference that $\vec{E}_i^{\;p}$ and $\vec{B}_i^{\;p}$ are calculated integrating the full electromagnetic fields over the shape function  $S(\vec{x}-\vec{x}_i^{\;p}(t))$ (see Eqs.~\eqref{eq:interpolation_em}-\eqref{eq:interpolation_em2}). The code actually solves the relativistic version of these equations, obtained by replacing $\vec{v}_i$ with the four-velocity $\vec{u}=\gamma\vec{v}_i$, with $\gamma=(1-v_i^2/c^2)^{-1/2}$ the relativistic gamma factor. But relativistic effects can be safely neglected for the regime appropriate for a quasineutral hybrid code ($v_i\ll c$, with $c$ the speed of light). The ion density $n_i$, ion bulk velocity $\vec{u}_i$ and the ion  current density $\vec{\jmath}_i$ are obtained as the first two order velocity momenta of the ion distribution function $f_i(\vec{x}_i,\vec{v}_i)$, i.e.,  $n_i=\int f_i\,d^3\vec{v}$ and $\vec{\jmath}_i=e\int \vec{v}_i\,f_i\,d^3\vec{v}=en_i\vec{u}_i$ (see details of the implementation in Sec.~\ref{sec:numerical_implementation}).  With a similar procedure it is possible to obtain higher order momenta such as the ion pressure tensor (associated with the ion temperature).

For fluid electrons, we consider the electron momentum equation with finite electron inertia and resistive effects, i.e., a generalized Ohm's law for the electric field:
\begin{align} \label{eq:e_ohm}
	\vec{E} & = -\vec{u}_e\times \vec{B} -\frac{1}{en_e}\frac{\partial P_{e,jk}}{\partial x_k} - \frac{m_e}{e}\left(\frac{\partial \vec{u}_e}{\partial t}+(\vec{u}_e\cdot\vec{\nabla})\vec{u}_e\right)  + \eta \vec{\jmath}.
\end{align}
Here, $\eta=m_e\nu/(e^2n_e) $ is the collisional resistivity with $\nu$ the collision frequency (between electrons and ions), $\vec{\jmath}=e(n_i\vec{u}_i - n_e\vec{u}_e)$ is the total current density, $\vec{u}_e$ is the electron fluid velocity and $P_{e,jk}$ is the $jk$ component of the electron pressure tensor. Note that $k$ is a summation index, in such a way that $\partial P_{e,jk}/\partial x_k$ represents the $j=x,y,z$ component.

The electric and magnetic fields are related to the plasma current via the Faraday's law and Amp\`ere's law without displacement current:
\begin{align}
	\vec{\nabla}\times \vec{E} & =-\frac{\partial \vec{B}}{\partial t}, \label{eq:faraday} \\
	\vec{\nabla}\times \vec{B} & =\mu_0\,n\,e(\vec{u}_i-\vec{u}_e). \label{eq:ampere}
\end{align}
Note that neglecting the displacement current in Eq.~\eqref{eq:ampere} is equivalent to set the dielectric constant $\epsilon_0\to 0$, preventing light  (and Langmuir) wave propagation and assuming quasi-neutrality $n_e=n_i=n$ for this parameter regime, with $n_{(e/i)}$ the electron/ion density, respectively. This condition implies the restriction to  frequencies much smaller than the electron plasma frequency \text{$\omega\ll\omega_{pe}$} and length-scales much larger than the Debye length $\lambda\gg \lambda_{De}$. The Poisson's equation $\vec{\nabla}\cdot\vec{E}=e(n_i-n_e)/\epsilon_0$ is automatically satisfied because of the quasi-neutral approximation and boundary conditions, while $\vec{\nabla}\cdot\vec{B}=0$ is also fulfilled if the initial conditions do so~\cite{Winske2003}.

Combining the electron momentum Eq.~\eqref{eq:e_ohm} and Faraday's law Eq.~\eqref{eq:faraday}, we can eliminate the electric field and obtain an evolution equation for the magnetic field, which has the form of a generalized continuity equation,
\begin{align} \label{eq:curl_emom}
	\frac{\partial \overrightarrow{W}}{\partial t}
	  & = \vec{\nabla}\times\left [\vec{u}_e\times \overrightarrow{W}\right]-\vec{\nabla}\times\left(\frac{\vec{\nabla} p_e}{m_en_e}\right)- \vec{\nabla}\times\left(\frac{\nu}{en_e}\vec{\jmath}\right),
\end{align}
where
\begin{equation}\label{eq:generalized_vorticity}
	\overrightarrow{W}=\vec{\nabla}\times\vec{u}_e-e\vec{B}/m_e,
\end{equation}
is the generalized vorticity.  For simplicity, we have assumed a scalar electron pressure $p_e$ given by the isothermal equation of state,
\begin{align} \label{eq:eos}
	p_e = \frac{1}{3}P_{e,kk} = n_ek_BT_e,
\end{align}
with $k_B$  the Boltzmann constant and $T_e$ the electron temperature, constant in time. Note that if the temperature $T_e$ and density $n_e$ are spatially uniform, the second term in Eq.~\eqref{eq:curl_emom} vanishes.

In conventional hybrid codes with electron inertia~\cite{Shay1998,Kuznetsova1998,Lipatov2002}, the electric field is obtained from Eq.~\eqref{eq:e_ohm} while the magnetic field is obtained from Eqs.~\eqref{eq:curl_emom} and \eqref{eq:generalized_vorticity}.
This is done, however, by neglecting some terms.
Substituting for  $\vec{u}_e$ from Eq.~\eqref{eq:ampere} and neglecting terms proportional to $\partial \vec{u}_i/\partial t$ and $\partial n_e/\partial t$ and electron-scale spatial variations of the density, the l.h.s. (left hand side) of Eq.~\eqref{eq:curl_emom} can be written entirely in terms of the time derivative of the magnetic field, i.e., as $\partial/\partial t[e(\vec{B}-d_e^2\vec{\nabla}^2\vec{B})/m_e]$, where $d_e=c/\omega_{pe}$ is the electron inertial length. The magnetic field is then obtained by solving an elliptic equation for $\vec{B}$.
Neglecting  $\partial \vec{u}_i/\partial t$ and $\partial n_e/\partial t$ in Eq.~\eqref{eq:curl_emom} is justified for a large mass ratio $m_i/m_e$. However, these approximations are not necessarily valid in all the situations of interest and need to be checked for each case, especially for simulations using an artificially low mass ratio  $m_i/m_e$.
The electric field is calculated directly from the electron momentum Eq.~\eqref{eq:e_ohm} by neglecting $\partial\vec{u}_e/\partial t$, which is clearly inconsistent with  keeping this term in Eq.~\eqref{eq:curl_emom}.

In our hybrid simulation model, we solve Eqs.~\eqref{eq:xp}-\eqref{eq:eos} without making any of these approximations.
Eq.~\eqref{eq:curl_emom} is solved for the generalized vorticity $\overrightarrow{W}$ and then, the magnetic field is calculated from an elliptic equation obtained
by combining Eq.~\eqref{eq:generalized_vorticity} with Amp\`ere's law  Eq.~\eqref{eq:ampere},
\begin{align}\label{eq:elliptic_b}
	\frac{1}{\mu_0e}\vec{\nabla}\times\left(\frac{\vec{\nabla}\times\vec{B}}{n_e}\right)+\frac{e\vec{B}}{m_e} & = \vec{\nabla}\times\vec{u}_i-\overrightarrow{W}.
\end{align}
Note that in Eq.~\eqref{eq:elliptic_b} we do not neglect the density variations at electron scales, and explicitly calculate the curl of the ion bulk velocity.
Next, the electric field is calculated from the electron momentum Eq.~\eqref{eq:e_ohm} (in the form of a generalized Ohm's law) by explicitly evaluating the time derivative term $\partial\vec{u}_e/\partial t$. For this purpose, we assume that ions quantities do not change much in a single time step, which is a small fraction of the electron gyroperiod. Thus, for a single time step ($\Delta t$)  of the simulation, $\vec{u}_e$ is obtained at $t_1=t_0+\Delta t$ and $t_2=t_0+2\,\Delta t$ by advancing Eq.~\eqref{eq:curl_emom} in two steps of length $\Delta t$ from $t_0$ to $t_2$.
This allows the calculation of  $\partial\vec{u}_e/\partial t$ at $t_1$ by a central finite difference scheme. In the second of the two time steps, the code reuses the same ion quantities $n_i$ and $\vec{\jmath}_i$ used in the first time step (with index $t_1$).

\subsection{Importance of the full electron inertial terms compared to other hybrid algorithms}

In general, any hybrid code with electron inertia
should be considered in scenarios with length scales
on the order of $d_e=c/\omega_{pe}$ the electron skin depth
and frequencies on the order of $\Omega_{ce}$.
But in order to demonstrate the importance of the additional electron inertial terms
of the generalized Ohm's law,
let us combine the relevant equations of our hybrid plasma model
(Eqs.~\eqref{eq:vp} and \eqref{eq:e_ohm}-\eqref{eq:ampere}),
in a way similar to the typically used in other hybrid codes,
without electron pressure and without resistivity $\eta=0$ for simplicity:

\begin{align}
	\vec{\nabla} \times \vec{E}' & = - \frac{\partial \vec{B}'}{\partial
	t},  \label{eq:E1}\\
	\vec{E}' & = \vec{E} - \frac{\partial}{\partial t} \left( d_e^2
	\vec{\nabla} \times \vec{B} \right)  \nonumber\\
	& = \frac{\vec{\jmath} \times \left( \vec{B} - d_e^2
	\vec{\nabla}^2 \vec{B} \right)}{n e} - \frac{\vec{\jmath}_i \times
	\vec{B}}{n e} \nonumber\\
	&\phantom{=}+ \alpha_1 \left[\frac{\vec{\jmath} \times d_e^2
	\vec{\nabla}^2 \vec{B}}{n e} - \frac{m_e}{e} \left(
	\vec{u}_e\cdot\vec{\nabla} \right) \vec{u}_e\right] - \frac{\alpha_2 m_e}{e}
	\frac{\partial
	\vec{u}_i}{\partial t} , \label{eq:E2}\\
	\vec{B'} & = \vec{B} - d_e^2 \vec{\nabla}^2 \vec{B} + \alpha_3
	\vec{\nabla} d_e^2 \times \left( \vec{\nabla} \times \vec{B}
	\right),  \label{eq:E3}\\
	m_i  \frac{d \vec{v}_i}{dt}   & = e \vec{E}
	+ e \vec{v}_i \times \vec{B} \nonumber\\
	& = e \left[ \vec{E}' + \alpha_4 \frac{\partial}{\partial t} \left(
	d_e^2 \vec{\nabla} \times \vec{B} \right) \right] + e \vec{v}_i \times
	\vec{B},  \label{eq:E4}
\end{align}
where in the last equation we have dropped the indices related to
the coarse-graining of particles and electromagnetic fields by the particle
shape functions (index ``p'').
Note that certain terms in the set of the equations
\eqref{eq:E1}-\eqref{eq:E4}
have been multiplied by parameters $\alpha_i$ ($i=1,\dots4$)
whose values can be chosen to be either unity or zero.
The full set of inertial terms equations (used in our code)
can be obtained by setting $\alpha_i=1$ ($i=1,\dots4$).
The first and third terms in Eq.~\eqref{eq:E2} are combined
to give $\vec{\jmath}\times\vec{B}/(ne)-\vec{\jmath}_i\times\vec{B}/(ne)
= \vec{\jmath}_e\times\vec{B}/(ne) = -\vec{u}_e\times\vec{B}$.
The second term in Eq.~\eqref{eq:E2} comes from one
of the terms resulting from the expansion of the
fourth term
$(m_e/e)(\vec{u}_e\cdot\vec{\nabla})\vec{u}_e$ in
the Ohm's law (Eq.~\eqref{eq:e_ohm}).
Since this term has been combined with the $\vec{j}\times B/(en)$ term
in Eq.~\eqref{eq:E2} (first parenthesis), we have
simultaneously subtracted with the second term in the square brackets,
so that for $\alpha_1=1$ we recover the Ohm's law in its usual form
(Eq.~\eqref{eq:e_ohm}).
We compare explicitly our equations with
the approximated set of hybrid plasma equations simulated by Ref.~\cite{Kuznetsova1998},
which
can be obtained by setting $\alpha_1=1$ and
$\alpha_2=\alpha_3=\alpha_4=0$,  and those simulated by \cite{Shay1998}
by setting  $\alpha_i=0$ ($i=1,\dots4$).

Now, let us estimate the orders of magnitude of each of the
commonly neglected terms related to the electron inertia in
Eqs.~\eqref{eq:E1}-\eqref{eq:E4}.

In order to estimate the importance of the terms with $\alpha_1$,
it is necessary to expand the second term $(m_e/e)(\vec{u}_e\cdot\vec{\nabla})\vec{u}_e$ in square brackets
by using vector identities and the Amp\`ere's law.
In Ref.~\cite{Kuznetsova1998}, the full terms with $\alpha_1=1$
are retained, while in Ref.~\cite{Shay1998} only part of the
aforementioned expanded second term is retained, neglecting
in particular  $(-m_e/e)\vec{\nabla}(\vec{u}_e^2)/2$.
Note that $\vec{\nabla}\vec{u}_e^2$, as part of the
electric field $\vec{E}'$ is always canceled
after substituting it in the curl $\nabla\times$
of the Faraday's law Eq.~\eqref{eq:E1}. But it
has to be considered in the ion equation of motion
Eq.~\eqref{eq:E4}, which is not done in Ref.~\cite{Shay1998}.
Then, approximating the scale of variation of the electron
fluid velocity as $\nabla u_e \sim u_e/L_u$, we can compare
$(-m_e/e)\vec{\nabla}(\vec{u}_e^2)/2$
with the combined form of the first and third term of Eq.~\eqref{eq:E2}, $-\vec{u}_e\times\vec{B}$,
\begin{align}
	\dfrac{
		\left|-\dfrac{m_e}{2e} \left(
	\vec{\nabla}\vec{u}_e^2 \right) \right|
	}{
		\left|\vec{u}_e\times\vec{B}\right|
	}
	\sim \dfrac{
		\dfrac{m_e u_e^2}{eL_u}
		}{
		u_eB
		}
	\sim \left(\frac{d_e}{L_u}\right)\left(\frac{u_e}{V_{Ae}}\right).
\end{align}
Therefore, the full term with $\alpha_1$ becomes important whenever
the length scale variation of the fluid velocity (shear)
is on the order of magnitude of the electron skin depth $d_e\sim L_u$
and/or the electron fluid velocity is on the order of
the electron Alfv\'en speed $u_e\sim V_{Ae}$ (which is very common
in magnetic reconnection, even at ion scales).

The term with $\alpha_1$ in Eq.~\eqref{eq:E2} can be compared
with the first term in the same equation. This yields:
\begin{align}
\dfrac{\left|
\dfrac{\vec{\jmath} \times d_e^2
	\vec{\nabla}^2 \vec{B}}{n e}
\right|}{
	\left|\dfrac{\vec{\jmath}\times\vec{B}}{ne}
	\right|
}
\sim \dfrac{\dfrac{jd_e^2 B}{L_B^2 ne}}{\dfrac{jB}{ne}}
\sim \left(\frac{d_e}{L_B}\right)^2.
\end{align}
i.e., it is not correct to neglect this term whenever the
length scale of the magnetic fields is similar to
the electron skin depth: $L_B\sim d_e$.
This term was considered by Ref.~\cite{Kuznetsova1998}
but not in Ref.~\cite{Shay1998}.

The term with $\alpha_2$ in Eq.~\eqref{eq:E2} can be also compared
with the combined form of the first and third terms of the same equation,
approximating $\partial/\partial t\sim\omega$, a typical fluctuation frequency:
\begin{align}
	\dfrac{
		\left|\dfrac{m_e}{e} \dfrac{\partial \vec{u}_i}{\partial t}\right|
	}{
		\left|\vec{u}_e\times\vec{B}\right|
	}
	\sim \dfrac{
		\dfrac{m_e u_i\omega}{e}
		}{
		u_eB
		}
	\sim \left(\frac{\omega}{\Omega_{ce}}\right)\left(\frac{u_i}{u_e}\right)
	\sim \sqrt{\frac{m_e}{m_i}} \left(\frac{\omega}{\Omega_{ce}}\right),
\end{align}
where we have assumed that the ratio between $u_i/u_e$ scales
(at least) as the inverse of the square root of the mass ratio, as in many
typical plasma speeds (e.g: Alfv\'en speed and the
electron Alfv\'en speed). This term, in general, will
be small, unless the frequencies are greater than the
electron cyclotron frequency: $\omega\gtrsim \Omega_{ce}$
and the mass ratio is small, as is often the case
in many plasma simulations where an artificial
mass ratio is used for computational convenience.
Note that this term was ignored in both Refs.~\cite{Kuznetsova1998,Shay1998}.

The term with $\alpha_3$ in Eq.~\eqref{eq:E3} can be
compared with the second term in that equation.
Since $d_e= {\rm constant}/\sqrt{n}$, the gradient
$\vec{\nabla} d_e^2 \sim -({\rm const}^2/n^2)\nabla n\sim
-{\rm const}^2/(nL_n)$, with $L_n$ the typical length scale of variation
of the density. Approximating $\nabla B \sim B/L_B$, the scale
length of variation of magnetic field, the estimation yields:
\begin{align}
	\dfrac{
		\left|\vec{\nabla} d_e^2 \times \left( \vec{\nabla} \times \vec{B}
	\right)\right|
	}{
		\left|d_e^2\vec{\nabla}^2\vec{B}\right|
	}
	\sim \dfrac{
		\dfrac{{\rm constant}^2\cdot B}{nL_n L_B}
		}{
		\dfrac{B\cdot {\rm constant}^2 }{n L_B^2}
		}
	\sim \left(\frac{L_B}{L_n}\right).
\end{align}
This term is relevant wherever the magnetic length scale gradient
is similar or greater than the density gradient: $L_B\gtrsim L_n$
(and both terms are, of course, on the order of $d_e$).
Such a case might take place during collisionless guide field
reconnection, in the cavities of the
low density separatrix~\cite{Pritchett2004}.
Note that this term was ignored in both Refs.~\cite{Kuznetsova1998,Shay1998}.

Finally, the term with $\alpha_4$ in Eq.~\eqref{eq:E4} can be compared
with the last term in the same equation.  The ratio between those two terms
yields:
\begin{align}
	\dfrac{\left|\dfrac{\partial}{\partial t}
	\left(d_e^2 \vec{\nabla} \times \vec{B} \right)\right|}{\left|\vec{u}_i\times\vec{B}\right|}\sim \frac{\omega d_e^2 B}{u_iBL_B}
= \dfrac{\omega}{\omega_{pe}}\frac{d_e}{L_B}\frac{c}{u_i}.
\end{align}
The first factor $\omega/\omega_{pe}$ is usually negligibly small,
but the last one $c/u_i$ can be very large.
So, in case the multiplication
of both terms leads to a factor of order 1, the term with $\alpha_4$
should not be neglected in case of steep gradients of
magnetic field, i.e.: $L_B\lesssim d_e$.
Note that $v_i$ is the microscopic velocity which
can be considered as part of the ion fluid velocity $u_i$ when
the ions are cold and the first-order fluid momenta are taken from Eq.~\eqref{eq:E4}.
Note that this term was ignored in both Refs.~\cite{Shay1998,Kuznetsova1998}.

\section{Numerical implementation \label{sec:numerical_implementation}}
\begin{figure}[!ht]
	\includegraphics[width=1.0\linewidth]{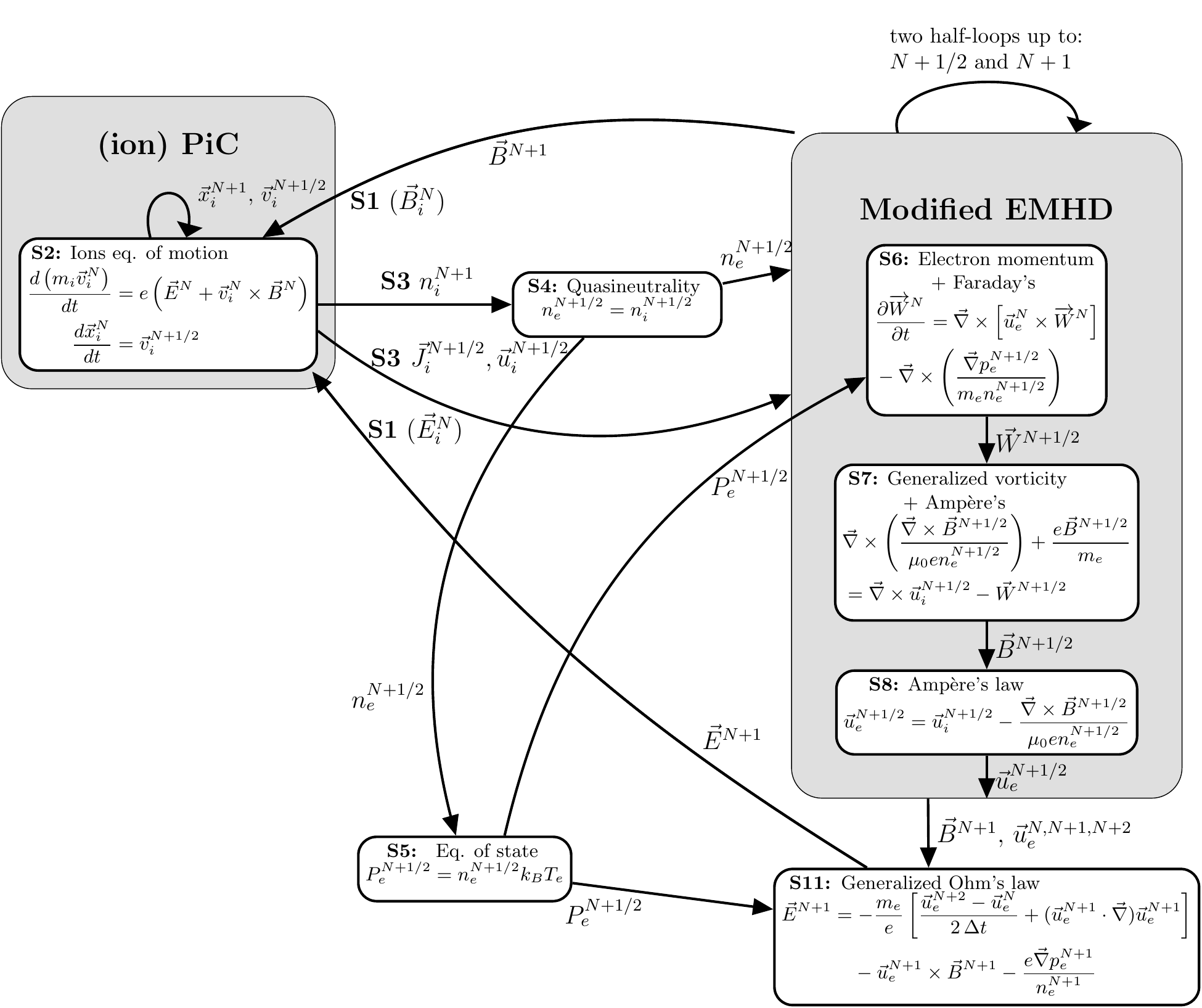}
	\caption{Scheme of the equations of our hybrid simulation code developed by coupling a PiC and an EMHD code. ``Sx'' stands for step number x (explained in the text). \label{fig:pic_emhd}}
\end{figure}

The simulation model discussed in Sec.~\ref{sec:model} is numerically implemented by combining the Particle-in-Cell (PiC) code ACRONYM~\cite{Kilian2012}
\footnote{
	\texttt{\url{http://plasma.nerd2nerd.org/}}
}, with an electron-magnetohydrodynamics (EMHD) code~\cite{Jain2003,Jain2006}. Both codes have been tested independently and used in the past for simulations of magnetic reconnection~\cite{Jain2014e,Munoz2016a}, instabilities~\cite{Jain2015c,Munoz2014a}, particle acceleration via instabilities~\cite{Burkart2010a}, CME-driven shocks~\cite{Kilian2015}, wave coupling~\cite{Ganse2014}, resonant wave-particle interaction~\cite{Schreiner2017}, etc.
Fig.~\ref{fig:pic_emhd} shows a scheme with the numerical implementation and coupling of the two codes.
This choice of algorithm was the fastest and safer way (less prone to numerical errors) to couple both codes without further modification and taking
advantage of their good properties.
But additional schemes  and algorithms to solve
those equations can be attempted in the future  to check
whether if they can provide an advantage to the existing ones.

On the PiC side of the code, the electromagnetic fields are assumed to be known at the staggered grid called the Yee lattice~\cite{Yee1966}. They are, however, provided by the EMHD side of the code on a grid where all quantities are defined at cell centers except the magnetic field which is defined at the edges of each cell. Therefore, we employ an interpolation procedure between the two grids schematized in Fig.~\ref{fig:interpolation}, in order to align correctly those quantities in the Yee lattice. Once the electromagnetic fields are interpolated from the EMHD to the PiC cell, the following steps (shown in Fig.~\ref{fig:pic_emhd}) are taken to move the ions:
\begin{figure}[!ht]
	\includegraphics[width=1.0\linewidth]{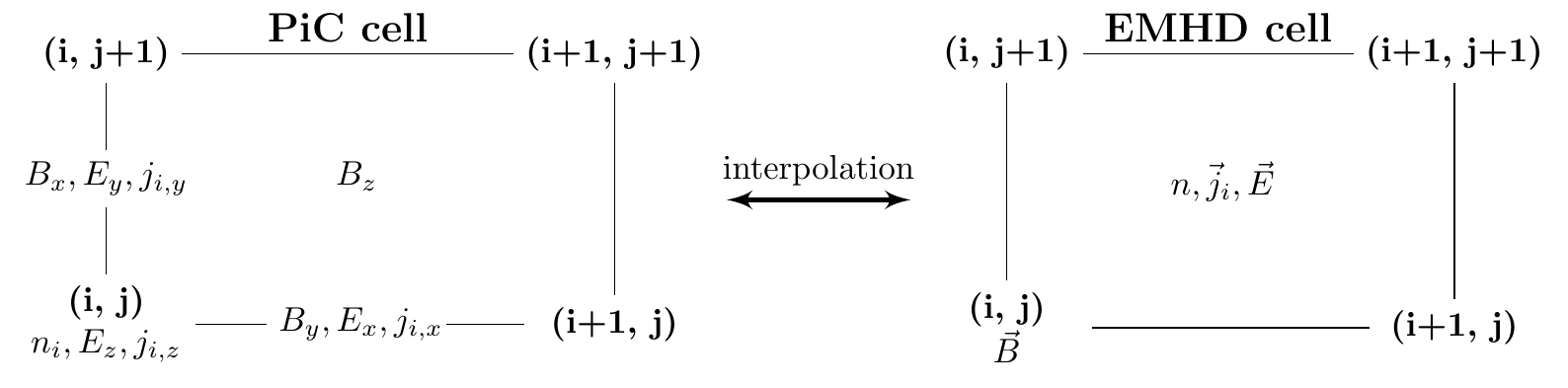}
	\caption{Scheme of the allocations of different physical quantities in the grids associated to the PiC (Yee lattice) and EMHD parts of the hybrid code. For simplicity, only the 2D case is shown. The indices of the grid points are indicated in the vertices of the cell ($i$ for the $x$ direction and $j$ for the $y$ direction). \label{fig:interpolation}}
\end{figure}
\subsection{Particle mover (PiC part)}\label{sec:pic_part}
The PiC side of the hybrid code, as done in all fully-kinetic PiC codes, including ACRONYM, updates ion velocities and positions in the given electric and magnetic fields $\vec{E}_i^N$ and $\vec{B}_i^N$ known at the time index $N$.
\begin{itemize}
	\item \textbf{Step 1:}
	The electromagnetic fields are interpolated from the grid to the macro-particle positions with a weighting given by the shape function $S(\vec{x}_i-\vec{x}_i^{\;p})$:
	\begin{align}\label{eq:interpolation_em}
		\vec{E}_i^{\;p}& = \vec{E}(\vec{x}_i^{\;p})=\int \vec{E}(\vec{x}_i)\,S(\vec{x}_i-\vec{x}_i^{\;p})\,d\vec{x}_i,                              \\
		\vec{B}_i^{\;p} &= \vec{E}(\vec{x}_i^{\;p}) = \int \vec{B}(\vec{x}_i)\,S(\vec{x}_i-\vec{x}_i^{\;p})\,d\vec{x}_i, \label{eq:interpolation_em2}
	\end{align}
	where the subscript $p$ indicates the location of each macro-particle.
	In the code, the interpolations between the EMHD and PiC grid and from the PiC grid to macro-particles' position  are combined in a single operation.
	The deposition of the particle quantities
	density $n_i$ and current density $j_z$ is not done to
	the grid points as in the usual Yee lattice, but to the grid centers as required by the EMHD solver.
	This is possible by centering  the particle shape functions
	around the grid center, and considering that
	there is no need to calculate these quantities in the PiC
	cell, since they are used as input to advance the electromagnetic
	fields in the EMHD side of the code, which requires only
	quantities living in the EMHD cell.
	This combination of interpolations has the advantage of less calculations
	that could introduce numerical errors otherwise, in addition
	to use the same identical scheme as the Step 3 below by construction.

	\item \textbf{Step 2:}
	Then, the ions can be moved via the second-order accurate leap-frog algorithm, which means that at each timestep $N$, the ion velocities are advanced from the  half-timestep $N-1/2$  to the half-timestep $N+1/2$ and the ion positions from the timestep $N$ to the timestep $N+1$ by using the discretized version of Eqs.~\eqref{eq:xp} and \eqref{eq:vp} (see, e.g., Sec.~4.3 of Ref.~\cite{Birdsall1991}):
	\begin{align}
		\label{eq:xp_num}  \vec{x}_i^{\;p,N+1}  & = \vec{x}_i^{\;p,N} + \vec{v}_i^{\;p,N+1/2}\Delta t ,                                                                                             \\
		\label{eq:vp_num} \vec{v}_i^{\;p,N+1/2} & = \vec{v}_i^{\;p,N-1/2} + \frac{e}{m_i}\left( \vec{E}_i^{\;p,N}+\frac{\vec{v}_i^{\;p,N+1/2}+\vec{v}_i^{\;p,N-1/2}}{2}\times \vec{B}_i^{\;p,N}\right)\Delta t.
	\end{align}
	Note that the first equation is explicit while the second implicit. We advance the ion velocity via the Boris method~\cite{Boris1970}, which involves a rotation of $\vec{v}_i^{\;N+1/2}$ (i.e., its magnitude is kept constant) making Eq.~\eqref{eq:vp_num} also explicit.

	\item \textbf{Step 3:}
	Because of the staggering of the position and velocity updates, the sources of the electromagnetic field, the ion number density $n_i$  and current density $\vec{\jmath}_i$, are computed at the time index $N+1$ and $N+1/2$, respectively.
	This deposition is done via an interpolation scheme using the same shape function  as for the electromagnetic field interpolation from the grid to the macro-particles position (Eqs.~\eqref{eq:interpolation_em}-\eqref{eq:interpolation_em2}):
	\begin{align}\label{eq:interpolation_n_j}
		n_i^{N+1}(\vec{x}_i) & = \sum_p N_p\,S(\vec{x}_i-\vec{x}_i^{\;p,N+1}) ,                                               \\
		\vec{\jmath}_i^{\;N+1/2}(\vec{x}_i) & = \sum_p N_p\,\vec{v}_i^{\;p,N+1/2} S(\vec{x}_i-\vec{x}_i^{\;p,N+1}).\label{eq:interpolation_n_j2}
	\end{align}
	The code has available, at compilation time, an option to smooth both quantities by performing a binomial filter and, if required, a compensation pass~\cite{Vay2011}. Note that, different from ACRONYM and other state-of-the-art fully-kinetic PiC codes, we do not need to use the charge-preserving Esirkepov scheme~\cite{Esirkepov2001} to deposit the current $\vec{\jmath}_i$  onto the grid, due to the quasineutrality condition of the hybrid approach\footnote{
	Other hybrid codes have used this method due to other reasons. For example, the hybrid code by Ref.~\cite{Swift2001} implemented  a charge preserving scheme~\cite{Villasenor1992} because it was computationally convenient for his curvilinear coordinate system.
	}.
	On the other hand, the reason to use the same shape function in Eqs.~\eqref{eq:interpolation_n_j} and \eqref{eq:interpolation_em} is to avoid self-forces and preserve the global momentum
	(see, e.g., Secs.~8.5-6 in Ref.~\cite{Birdsall1991}, Sec 5.3.3 in Ref~\cite{Hockney1988} or Ref.~\cite{Brackbill2016} for fully-kinetic PiC codes, or Sec.~4.5.2 in Ref.~\cite{Lipatov2002} for hybrid-PiC codes).
	Our hybrid code has available, as inherited from ACRONYM, the following shape functions:
	NGP (Nearest Grid Point), CIC (Cloud in Cell), TSC (Triangular Shaped Cloud), PQS (Piecewise Quadratic Spline) and other ones, useful under some specific conditions. By default, we use the second order shape function TSC to reduce the intrinsic PiC shot noise and
	the associated numerical heating without affecting the computational performance too much~\cite{Cormier-Michel2008,Kilian2013b,Munoz2014a}.
	Once the sources of the electromagnetic fields in Eqs.~\eqref{eq:interpolation_n_j}-\eqref{eq:interpolation_n_j2} are known at the Yee lattice, they are passed and interpolated to the EMHD grid following the inverse procedure as in Fig.~\ref{fig:interpolation} (similarly as Step 1, the operation is actually combined with the deposition in Eqs.~\eqref{eq:interpolation_n_j}-\eqref{eq:interpolation_n_j2}).
	Then, the EMHD part of the code updates the electromagnetic fields by using the electron fluid equations  and Maxwell's equations described later. The updated electric and magnetic fields are then passed back to the PiC side and the full cycle is repeated.

\end{itemize}
\subsection{Field updater (EMHD part)}\label{sec:emhd_part}
After step 3 in Fig.~\ref{fig:pic_emhd}, just before entering into the EMHD part of the code, the input quantities are known at the following time indices: $\vec{E}^N$, $\vec{B}^N$, $\vec{\jmath}^{N+1/2}$ and
$n_i^{N+1}$.
The ion density is also calculated at the same index, $n_i^{N+1/2}$, by taking the average value of $n_i^{N}$ and $n_i^{N+1}$.

\begin{itemize}

	\item \textbf{Step 4:}
	The quasineutrality condition is applied to get the electron density $n_e^{N+1/2}=n_i^{N+1/2}$.

	\item \textbf{Step 5:}
	With $n_e^{N+1/2}$ and a specified value of the electron temperature  $T_e$ (which is constant and  does not evolve in time), the code computes the electron pressure $p_e^{N+1/2}=n_e^{N+1/2}k_BT_e$ via the equation of state, Eq.~\eqref{eq:eos}.

	\item \textbf{Step 6 (first half-loop):}
	In order to update the electric and magnetic fields, the generalized vorticity  $\overrightarrow{W}$ in Eq.~\eqref{eq:curl_emom} is advanced from the timestep $N$ to $N+1$ by using the flux-corrected transport algorithm~\cite{Boris1993} of the LCPFCT package
	\footnote{
		\texttt{\url{http://www.nrl.navy.mil/lcp/LCPFCT}}
	},
	which is specially convenient to resolve steep gradients,
	one of the main applications of our code.
	The full update to the next timestep ($\overrightarrow{W}^{N+1}$) requires  $\overrightarrow{W}^{N+1/2}$ and $\vec{u}_e^{N+1/2}$, which are not known yet. Because of this, the field solver first advances the  generalized vorticity $\vec{W}^N$ by half a timestep (with the input quantities  $n_e^{N+1/2}$, $p_e^{N+1/2}$, $\vec{u}_e^{N}$,  $\vec{W}^N$, $\vec{B}^N$) to estimate $\vec{u}_e^{N+1/2}$,  $\vec{W}^{N+1/2}$, and $\vec{B}^{N+1/2}$,  by solving the discretized version of Eq.~\eqref{eq:curl_emom}:
	\begin{align}\label{eq:curl_emom_discrete}
		\left.\frac{\partial \overrightarrow{W}}{\partial t}\right|_{N\to N+1/2}
		  & = \vec{\nabla}\times\left [\vec{u}_e^N\times \overrightarrow{W}^N\right]-\vec{\nabla}\times\left(\frac{\vec{\nabla} p_e^{N+1/2}}{m_en_e^{N+1/2}}\right).
	\end{align}
	A multi-dimensional time operator splitting technique
	(see, e.g., Sec.~5.3.1 of Ref.~\cite{Lipatov2002} or Sec.~20.3.3 of Ref.~\cite{Press2007})
	is used to solve this equation
	by decomposing it into six sub-equations, grouped
	in three sets (one for each direction of integration $x$, $y$, $z$) composed
	of terms containing the spatial derivative along the integration direction and a time derivative
	of the generalized vorticity.
	Then,  step 6 to step 8 are performed for each direction of integration.

	\item \textbf{Step 7 (first half-loop):}
	Next, the solution for $\vec{B}^{N+1/2}$ of the discretized version of the elliptic Eq.~\eqref{eq:elliptic_b} is obtained,
	\begin{align}
		\vec{\nabla}\times\left(\frac{\vec{\nabla}\times\vec{B}^{N+1/2}}{\mu_0en^{N+1/2}}\right)+\frac{e\vec{B}^{N+1/2}}{m_e} & =\vec{\nabla}\times\vec{u}_i^{N+1/2}-\vec{W}^{N+1/2}\label{eq:b_half}.
	\end{align}
	Its solution allows to obtain $\vec{B}^{N+1/2}$ from $\vec{W}^{N+1/2}$. This elliptic partial differential equation (PDE) is solved by the subroutines provided by the MUDPACK 5.0 libraries~\footnote{
		\texttt{\url{https://www2.cisl.ucar.edu/resources/legacy/mudpack}}
	},
	which use the multigrid iteration method~\cite{Adams1989,Adams1991}.
	In particular, the form of the equation is appropriate to be solved via the ``mud2/3'' subroutine, which is a second-order difference approximation for non-separable PDEs on a rectangular/cubic domain (for 2D/3D cases, respectively).

	\item \textbf{Step 8 (first half-loop):}
	The next step consists in obtaining the components of $\vec{u}_e^{N+1/2}$ from the discretized form of Amp\`ere's law Eq.~\eqref{eq:ampere}:
	\begin{align}
		\vec{u}_e^{N+1/2}=\vec{u}_i^{N+1/2}-\frac{\vec{\nabla}\times\vec{B}^{N+1/2}}{\mu_0en_e^{N+1/2}}\label{eq:ue_half}.
	\end{align}

	\item \textbf{Step 9: Steps 6-8, second half-loop:}
	At this point of time, the code knows all relevant quantities at the time index $N+1/2$, i.e., $\vec{u}_e^{N+1/2}$, $\vec{B}^{N+1/2}$ and $\vec{W}^{N+1/2}$.
	Then, all these quantities are advanced to the time index $N+1$, by repeating the steps 6 to 8 in Fig.~\ref{fig:pic_emhd}. Explicitly, the equations to be solved in this second half-loop are:
	\begin{align}
		\left.\frac{\partial \overrightarrow{W}}{\partial t}\right|_{N+1/2\to N+1}
		& = \vec{\nabla}\times\left [\vec{u}_e^{N+1/2}\times \overrightarrow{W}^{N+1/2}\right]-\vec{\nabla}\times\left(\frac{\vec{\nabla} p_e^{N+1/2}}{m_en_e^{N+1/2}}\right), \\
		\vec{\nabla}\times\left(\frac{\vec{\nabla}\times\vec{B}^{N+1}}{\mu_0en_e^{N+1/2}}\right)+\frac{e\vec{B}^{N+1}}{m_e} & =\vec{\nabla}\times\vec{u}_i^{N+1/2}-\vec{W}^{N+1}\label{eq:b_full},                                                                                                 \\
		\vec{u}_e^{N+1}                                                                                                     & =\vec{u}_i^{N+1/2}-\frac{\vec{\nabla}\times\vec{B}^{N+1}}{\mu_0en_e^{N+1/2}}\label{eq:ue_half_full}.
	\end{align}
	With this, we have the quantities $\vec{u}_e^{N+1}$, $\vec{B}^{N+1}$ and $\vec{W}^{N+1}$. Note that we have used the ion quantities ($\vec{u}_i$ and $n_i$) at the time index $(N+1/2)$, same as for the first half-loop. This is justified by the heavy mass of the ions compared to the electrons.

	The two half-loops described before (steps 6 to 8) are repeated for the integration along each one of the spatial directions $x,y$ and $z$ (in the 3D case), as required
	by the multi-dimensional time operator splitting technique in step 6, thus advancing
	in time all the components of the quantities used by field solver.

	\item \textbf{Steps 10: Steps 6-9, full-loop, second time:}
	Before calculating the electric field, we repeat the above procedure (Steps 6-9) by advancing the equations from time step $N+1$ to $N+2$ but also using the ion quantities at the time index $N+1/2$ (i.e., assuming that they are temporally fixed). This gives us  $\vec{B}^{N+2}$, $\vec{W}^{N+2}$ and $\vec{u}_e^{N+2}$.

	\item \textbf{Step 11:}
	Now, the electric field at the time step $N+1$ is calculated from the discretized form of the generalized Ohm's law Eq.~\eqref{eq:e_ohm}:
	\begin{align}
		\vec{E}^{N+1}=-\frac{m_e}{e}
		\left[
		\frac{\vec{u}_e^{N+2}-\vec{u}_e^{N}}{2\, \Delta t}
		+(\vec{u}_e^{N+1}\cdot\vec{\nabla})\vec{u}_e^{N+1}
		\right]
		-\vec{u}_e^{N+1}\times\vec{B}^{N+1}
		-\frac{\vec{\nabla} p_e^{N+1}}{en_e^{N+1}}.
	\end{align}
	We use the value of $p_e^{N+1}=n_e^{N+1}k_BT_e$ provided by the equation of state Eq.~\eqref{eq:eos} in the same way as before (steps 4 and 5 in Fig.~\ref{fig:pic_emhd}). Note that we have assumed $p_e^{N+1}=p_e^{N+1/2}$ and $n_e^{N+1}=n_e^{N+1/2}$.

	Finally, the values of $\vec{E}^{N+1}$ and $\vec{B}^{N+1}$ are passed to the PiC side of the hybrid code,  which now provides $\vec{\jmath}_i^{N+3/2}$ and $n_i^{N+2}$. This ends a timestep and the cycle is repeated for the following one.

\end{itemize}

The algorithm used in the code is second order accurate in space, as a result of the previous discretization of the equations.  A measurement of this error, with ion effects included,
is hindered because the PiC shot noise affects
any measure of error below the noise level, and even
the initial conditions will show some fluctuations
at this level. But without ion effects (only fluid quantities)
 a comparison of the error with analytical solutions
can be performed with higher accuracy showing the second order convergence in space.
This is proven numerically in the first of our test problems (see Sec.~\ref{sec:numerical_normal_modes}).

\subsection{Boundary conditions}
The EMHD Maxwell solver of the code relies on specific
libraries to solve some of the equations.
Both of them, LCPFCT and MUDPACK, allows to specify periodic boundary
conditions or a given value or its derivative to the input variables.
That easiness is the reason because periodic and
reflecting/PEC (perfect electric conductor)
are currently  implemented in the code.
Other more complicated boundary conditions such
as absorbing or open (already implemented in the ACRONYM PiC-code)
would require a modification of the EMHD Maxwell solver.

\subsection{Numerical stability}

The time advancing part of our hybrid code is solved
by the flux corrected transport method, and mostly their properties control
the stability of the overall algorithm.
A local von Neumann stability analysis of
this scheme
results in an expression for the amplification factor $|A|=|W^{N+1}/W^N|$,
whose specific form varies for the specific coefficients of diffusion and
anti-diffusion used in the algorithm~(see Refs.~\cite{Boris1973,Book1975,Boris1976}).
The stability condition (strictly for $W$, but it can also be applied
to $\vec{B}$ due to the second order discretization in space
of the elliptic equation~\eqref{eq:elliptic_b})
can then be calculated as $|A|\leq 1$,
and most of the different versions lead to the same local CFL condition:
$|\epsilon| =|V_e\Delta t/\Delta x|\leq 1$,
where $V_e$ is a typical (electron) speed in the system.
As explained below, and based on the maximum speed of the whistler
wave for this system, $V_e$ can be taken in many cases as the electron Alfv\'en speed.
Of course, some specific simulated plasmas can develop high electron
fluid velocities, in which case $V_e$ would correspond to the
maximum of those electron velocities in the system.
This (linear and local) requirement is, however, not strict: the flux
corrected transport algorithm
can be stable with larger timesteps, with an upper bound that is usually
found empirically.
This is due to the non-linear flux
correction (suppressing spurious short-wavelength maxima/minima caused
by numerical instabilities) and some non-local properties of the algorithm (since it
uses a stencil of more than 7 points in space).
For a specific hydrodynamic case, this scheme was found (by means
of numerical tests) to be  stable
for CFL numbers $|V\Delta t/\Delta x|\sim 1.2-1.3$~\cite{Boris1973}.

A basic plasma wave mode present in all magnetized plasmas is the whistler wave.
That is why it is convenient to use it as a basis for the
calculation of a CFL condition and choice of the time step.
Without ion effects and assuming a constant density, the system
of equations \eqref{eq:e_ohm}-\eqref{eq:ampere}
 (in $\vec{W}$, $\vec{u}_e$ and $\vec{B}$)
simplifies significantly.
Linearizing
around an equilibrium with uniform magnetic field, the resulting
whistler dispersion relation becomes~\cite{Bulanov1992},
	\begin{align}\label{eq:whistler}
		\frac{\omega}{\Omega_{ce}}=\frac{(kk_{\parallel})(d_e)^2}{1+(kd_e)^2},
	\end{align}
where $d_e$ is the electron skin depth.
From here it can be clearly seen
that in the limit without electron inertia $d_e\to0$ (as
in most of the hybrid codes), the denominator
in the r.h.s. becomes 1 and the frequency increases quadratically with
the wavenumber $k$,
implying an increasing phase speed without bounds for short
enough wavelengths. Therefore, those codes will become unstable
unless a very small time step is chosen, or other (numerical) diffusive
term is added in their algorithm.
For a finite $d_e$, the frequency of the whistler branch
increases in a similar way proportional to $k^2$ for relatively small frequencies
$\Omega_{ci}<\omega\ll\Omega_{ce}$.
But for higher frequencies, $\omega$ reaches an asymptote
at $\Omega_{ce}$ with a zero phase speed.
This makes the algorithm automatically stable against whistler wave
propagation, as long as the maximum wave speed
($V_{Ae}/2=\Omega_{ce}d_e/2$ reached at $k=d_e^{-1}$) is considered by the
CFL condition mentioned above.
Another important constraint is that the ion gyration should be
well resolved, leading to $\Delta t\Omega_{ci}\lesssim 0.3$.
Finally, it is worth to mention that in cases where all the
important phenomena take place uniquely at ion time and spatial scales,
electron inertia can be switched off in the code, with the
consequent possibility to choose a larger timestep,
on the order of the (ion) Alfv\'en speed
$V_{A}\Delta t/\Delta x\lesssim 1$.
But note that, in this case, all the electron physics and frequencies
above $\Omega_{ci}$ are severely modified.

\section{Test problems \label{sec:tests}}

In the following we describe one numerical and six physical test problems used to validate our numerical method.

\begin{table}[!ht]
		\begin{tabular}{llcccc}
			Test problem             & $m_i/m_e$ & $V_A/c$   & $\beta_i$       & $T_i/T_e$  \\ \hline
			(0) Excitation of parallel eigenmodes   & $100$     & $10^{-4}$ & $0.1$       & 1          \\
			(1) Parallel EM modes        & $100$     & $10^{-4}$ & $\{0.01,0.1\}$       & 1          \\
			(2) Ion Bernstein modes            & $400$     & $10^{-3}$ & 0.1             & 1          \\
			(3) Ion beam R instability   & $100$     & $10^{-4}$ & $\beta_e$=0.1 & 10         \\
			(4) Ion Landau damping       & $100$     & $10^{-3}$ & 2.0             & [0.1-0.66] \\
			(5) Ion firehose instability & $100$     & $10^{-2}$ & $300/\pi=95.5$       & 1\\
			(6) 2D oblique ion firehose instability & $100$     & $10^{-3}$ & $2.8$       & 1
		\end{tabular}
	\caption{Main physical parameters of the tests problems. The ion Landau problem is unmagnetized, but we still use a reference magnetic field for calculating $\beta_i$ and all the normalizations depending on it. We use two values for $\beta_i$  in the parallel EM modes, and a range of values for $T_i/T_e$ for the ion Landau damping test problem. Since there are different values of $\beta_i$ for the different components of the ion beam R instability, here we  give instead $\beta_e$. \label{tab:physical_parameters_tests}}
\end{table}

\begin{table}[!ht]
		\begin{tabular}{ccccccc}
			Test problem             & $N_x$ & $L_x/d_i$ & $\Delta x/d_i$ & $\Delta t\,\Omega_{ci}$ & $T\Omega_{ci}$ & ppc  \\ \hline
			(1)        & 1024  & 102.4     & $0.1$          & $5\cdot10^{-3}$       & 600                 & 512  \\
			(2)            & 1920  & 63.6      & $0.033$        & $4\cdot10^{-3}$       & 600                 & 64   \\
			(3)   & 1024  & 256       & $0.25$         & $1\cdot10^{-2}$       & 120                 & 1024 \\
			(4)       & 256   & 42.24     & $0.165$        & $1\cdot10^{-3}$       & 60                  & 4096 \\
			(5)  & 480   & 300       & $0.625$        & $1\cdot10^{-2}$       & 2400                & 2048 \\
			(6)  & $256$   & $256$       & $1.0$        & $0.8\cdot10^{-2}$       & 400                & 256 \\
			 & $N_y=128$   & $L_y/d_i=128$       &         &      &                 &
		\end{tabular}
	\caption{Main numerical parameters of the test problems.
	$N_x$ is the number of grid points along the resolved direction,
	$L_x$ is the simulation box length,
	$\Delta x$ is the grid resolution,
	$\Delta t$ is the time-step,
	$T$ is the total simulation time
	and ppc are the number of particles per cell.
	We use $N_y=4$ grid points in the transverse direction for all 1-D runs, while the grid size for the 2D firehose test problem is explicitly indicated.
	The ion Landau problem is unmagnetized, but we still use a reference magnetic field for calculating $\Omega_{ci}$ and all the normalizations depending on it.
	The numerical parameters of the first numerical eigenmodes test problem (0) are not shown
	because all of them are varied.
\label{tab:numerical_parameters_tests}}
\end{table}

\subsection{Excitation of parallel eigenmodes \label{sec:numerical_normal_modes}}

In order to test the accuracy of our algorithm, we carried out
several sets of simulations based on
the parallel propagating normal modes.
In the framework of the cold two-fluid plasma model, the two basic parallel (to a background magnetic field) propagating electromagnetic waves are the L and R modes. Their name stand for their polarization: L/R for left/right-handed circularly polarized waves, sometimes called ion-cyclotron Alfv\'en and whistler waves, respectively. These waves have electric and magnetic field fluctuations perpendicular to their propagation direction $\vec{k}$. Their dispersion relations are given by (see, e.g., Eqs.~6.49-50 of Ref.~\cite{Boyd2003} or Eq.~2.5 of Ref.~\cite{Stix1992}):
\begin{align}\label{eq:lr_mode}
	\left(\frac{ck}{\omega}\right)^2
	& = 1 - \frac{\omega_{pe}^2}{\omega}\frac{1}{\omega \pm \Omega_{ce}} - \frac{\omega_{pi}^2}{\omega}\frac{1}{\omega \mp \Omega_{ci} },
\end{align}
where the signs in the denominator of the first ($\pm$) and second term  ($\mp$)  of the right hand side correspond to L/R modes, respectively. Here, $\omega_{p\{e/i\}}$ are the electron/ion plasma frequencies and $\Omega_{c\{e/i\}}$ are the electron/ion cyclotron frequencies. In order to compare our results with the models used in the classical hybrid codes with massless electrons, we use the approximation of Eq.~\eqref{eq:lr_mode} for low frequencies and $m_e\to0$:
\begin{align}\label{eq:lr_lowfreq}
	\qquad (kd_i)^2 & = \frac{(\omega/\Omega_{ci})^2}{1 \pm \omega/\Omega_{ci}},
\end{align}
where the $\pm$ signs correspond to the R/L modes (see also the Eq.~2.45 in Ref.~\cite{Cramer2001}).
Note that we have used the relations $V_{A}=d_i\Omega_{ci}$ and $V_{A}/c=\Omega_{ci}/\omega_{pi}$, where  $V_A$ is the Alfv\'en speed and $d_i$ is the ion skin depth.
For higher frequencies around $\omega<\Omega_{ce}\ll\omega_{pe}$, the R
mode is usually called whistler wave, with a dispersion relation
given by Eq.~\ref{eq:whistler}.


The initial setup is the following for a parallel
propagating wave along the $x$ direction with a background and constant
magnetic field $\vec{B}=B_0\hat{x}$.
A perturbation of the magnetic field in the form:
\begin{align}\label{initial_wave}
	\delta B_y(x) & =\phantom{\pm}\delta B\cos(k_0x - \omega_0 t ),\\
	\delta B_z(x) & = \pm \delta B\sin(k_0x - \omega_0 t ),
\end{align}
(for $t=0$, and so only $k_0$ is initially chosen, not $\omega_0$ which
is selected by the simulated plasma),
where the signs $\pm$ are chosen to take into account the polarization of
the left (L) and right (R, whistler) handed waves, respectively
(and thus to have in each case forward propagating waves).
The perturbation strength is chosen as $\delta B=0.05B_0$, in order
to be substantially above the numerical noise level, but low
enough in order to avoid non-linear effects such as harmonics and parametric
decay into other waves~\cite{Terasawa1986}.
The density is kept constant, while other physical parameters are
summarized in Tables~\ref{tab:physical_parameters_tests}.
Boundary conditions are periodic and the tests are
quasi-1D (mostly along one direction, averaging over 4 cells in the transverse direction).

The tests use a different set of numerical parameters in order to isolate
the individual effects to be analyzed, classified as follows:

\subsubsection{\label{dx_whistler} Convergence on $\Delta x$ of a whistler wave: Accuracy of the EMHD solver.}

In order to isolate the effects of the EMHD solver from ion kinetic effects,
we ran these tests without ion response, i.e., with density constant and
the ion current always zero. Both ions and electrons are also cold $\beta_e=\beta_i=0$.

We fit four wavelengths of a whistler wave (R mode) with $kd_i=1$ in
the simulation box (the wavelength is kept fixed and equal to $\lambda=2\pi d_i$).
We resolve each wavelength by a variable number of grid points in the range
$[8-2048]$, with a corresponding grid cell size $\Delta x/d_i = [0.7854-0.00306]$
($\Delta x/d_e = [7.854-0.0306]$).
The timestep is chosen to fulfill the CFL condition for the electron
Alfv\'en speed at the level $V_{Ae}\Delta t/\Delta x = 0.5$,
and so is variable between $\Delta t\Omega_{ci,B0}=[0.04-1.5\times10^{-4}]$.
Note that this restrictive condition is strictly not needed for this kind of relatively
low frequency wave, since their phase speed does not approach
to $V_{Ae}$, but it is only chosen to keep consistency with
later investigations (which develop high frequency waves).

We let the wave travel for four periods $T=2\pi/\omega$, as calculated from the
theoretical dispersion relation (see Eq.~\eqref{eq:lr_mode}).
First, we compare the wave shape at this point in time with the
analytical solution Eq.~\ref{initial_wave}, which has only one
free parameter to adjust, the phase $\omega_0 t$. This is because we
are assuming that the wave does not damp out (so $\delta B$ is constant),
strictly true for all the R
branch and the low frequency part ($k\lesssim 0.5$)
of the L mode (see Fig.~\ref{fig:l_modes_zoom}), and also that the initial
wave number $k_0$ is not modified (no mode conversion).
The comparison of wave shapes is done with the so-called L1-norm,
\begin{align}\label{error}
\epsilon = \frac{1}{ N \delta B}\sum_{i=1}^{N}|B_{y,simulation}(x_i) - \delta B_{y,analytical}(x_i,t)|,
\end{align}
i.e, the (absolute value of the) average difference between the
simulation and the analytical profile (normalized in such a way that
an error equal to the wave amplitude is 1).

Fig.~\ref{fig:n_convergence_whistler}a) (blue line) shows the results, showing a decreasing
error with increasing resolution of the wave shape.
A straight line $N^{-2}$ fits very well the simulation results, proving
the second order accuracy of the underlying hybrid algorithm,
consequence of the discretization by second order finite differences
of the corresponding equations.
However, beyond $N=1024$ the convergence degrades: the increasing
resolution actually increases the error.
We checked that for all the cases with $N\leq1024$ the maximum error
is located in the steepest gradients region (close
to $x/\lambda=\pi/2$ and $x/\lambda=3\pi/2$), but it becomes more homogeneously distributed
for $N=2048$ and beyond, indicating a different source of error not clearly
identified so far. We will discuss some possible reasons for this behavior later in this section.

\subsubsection{Time step $\Delta t$ effects on a whistler wave}

\begin{figure}[!ht]\centering
	\includegraphics[width=0.99\linewidth]{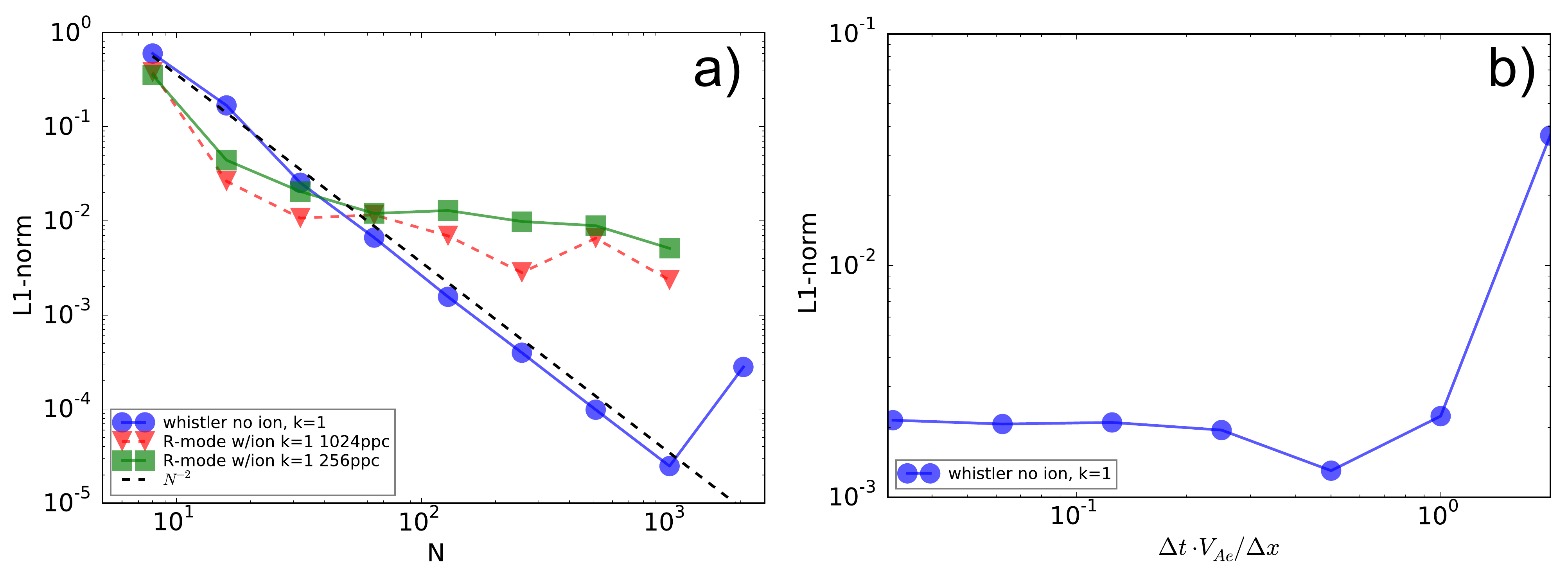}
	\caption{Convergence of a whistler wave with (red) and without (blue) ion effects for $kd_i=1$ on a) Number of points N per wavelength (inversely proportional to $\Delta x$) and b) time step $\Delta t$.
	\label{fig:n_convergence_whistler}}
\end{figure}

In order to test the effects of the time step on the convergence of the solution,
we use the same setup as the previous point for $N=64$ varying only
$\Delta t\Omega_{ci}=[3\times10^{-4}-0.02]$, in such a way that the CFL condition for the
electron Alfv\'en speed varies between
$V_{Ae}\Delta t/\Delta x=[0.003-2.0]$. The results in Fig.~\ref{fig:n_convergence_whistler}b) show that for a small enough
time step there is practically no improvement on the accuracy for
further decreasing values.
Only after the CFL condition is above the level $1.0$,
the accuracy decreases. The algorithm is still stable for
$V_{Ae}\Delta t/\Delta x=2.0$, but becoming unstable for values
equal or above 4.0.
%
Note that the code can run stably for values much higher than
$V_{Ae}\Delta t/\Delta x=1$ when the waves in the physical
system to be simulated have typically much lower frequencies
$\omega \ll \Omega_{ci}$, as it is the case for most of the test
problems analyzed here
(for which a CFL condition on the ion Alfv\'en speed is enough).

\subsubsection{\label{item2_ion} Ion effects (including particle number) on the convergence of a whistler wave}

Let us repeat the same previous test adding the full ion effects,
which requires also to  initialize the velocity field in order to
have a stable initial equilibrium.
For Alfv\'en waves, that magnetic perturbation
is associated to an ion velocity perturbations by means of the so-called
Wal\'en relation,
\begin{align}\label{walen}
	\delta \vec{V}_{\perp} = -(k_0/\omega_0)\delta \vec{B}_{\perp},
\end{align}
where $(\omega_0/k_0)=V_A$ is the (ion) Alfv\'en speed, characterizing their
non dispersive nature (group and phase speed are identical.
See, e.g., Sec. $\S$ 69. of Ref.~\cite{Landau1960}).
For the dispersive L and R (whistler) waves, where the Hall term
introduces dispersion, the velocity perturbation is also identical,
replacing $\omega_0$ by the value given by the appropriate dispersion relation~\cite{Terasawa1986}.
Note that this is not only valid for cold plasmas, but also
for hot plasmas, where full kinetic effects are taken into account~\cite{Sonnerup1967a}.
Other physical parameters to be considered are a constant temperature,
with $\beta_i=0.1$ (relatively high to emphasize the thermal
effects), and $T_i=T_e$ (to avoid unstable sound waves in the
regime $T_e\gg T_i$). We used 1024 particles per cell in this test.

Fig.~\ref{fig:n_convergence_whistler}a) (red line) shows the results
of the convergence study varying the number of points per wavelength
in the same way as in the item~\ref{dx_whistler} without ion effects.
For low resolution, there is a tendency to follow the same scaling
$N^{-2}$ of pure electron whistler waves, but then there is saturation
and the error does not decrease significantly for $N\gtrsim 64-128$,
keeping stable at the level of $0.5\%$ (on average).
For $N=1024$, the error with ion effects (red line) is two orders of magnitude larger
than the one without ion effects (blue line).
Different from the case without ion effects, in this case
the distribution of numerical error in the wave shape is more random,
and not located mostly in the points with strong gradients.
This is consequence of the numerical fluctuations due to the finite
number of macroparticles.


There are several explanations for this lack of good convergence of the L1-norm.
First of all, note that there are very few other hybrid-PiC or fully-PiC
based algorithms showing this specific error measure,
perhaps due to the sensitivity to the particle shot noise.
To our best knowledge, only one other hybrid-PiC algorithm~\cite{Kunz2014}
has reported those results, although using a $\delta f$ method (less noisy).
So, it is hard to compare how accurate is our algorithm with
other similar kind of codes.
Nevertheless, the saturation and further increase of this error has been seen
in other Lagrangian, particle-based simulations, although
for smooth particle hydrodynamic codes (SPH~\cite{Hopkins2015}).
It was argued that this effect is due to the particle shot noise and
the associated lack of equilibrium of the
initial setup: in our case, the initial ion velocity field does not match
the analytical profile regardless the spatial resolution (it is controlled
by the particle number associated to the random particle motion),
and thus, the error cannot decrease further than this initial error.
In order to check that the numerical thermal fluctuations are the reason
for this limit in the accuracy of the algorithm, we proceed
to repeat this test with fewer particles per cell, 256.
This choice increases twice the numerical noise.
The results
can be seen in the line green of Fig.~\ref{fig:n_convergence_whistler}a).
The numerical error is in general somewhat larger than the case with 4 times
more particles (red line) and slowly decreasing with increasing spatial
resolution, but saturating in a similar way as before.

\subsubsection{Convergence on $\Delta x$ for other waves}

\begin{figure}[!ht]\centering
	\includegraphics[width=0.99\linewidth]{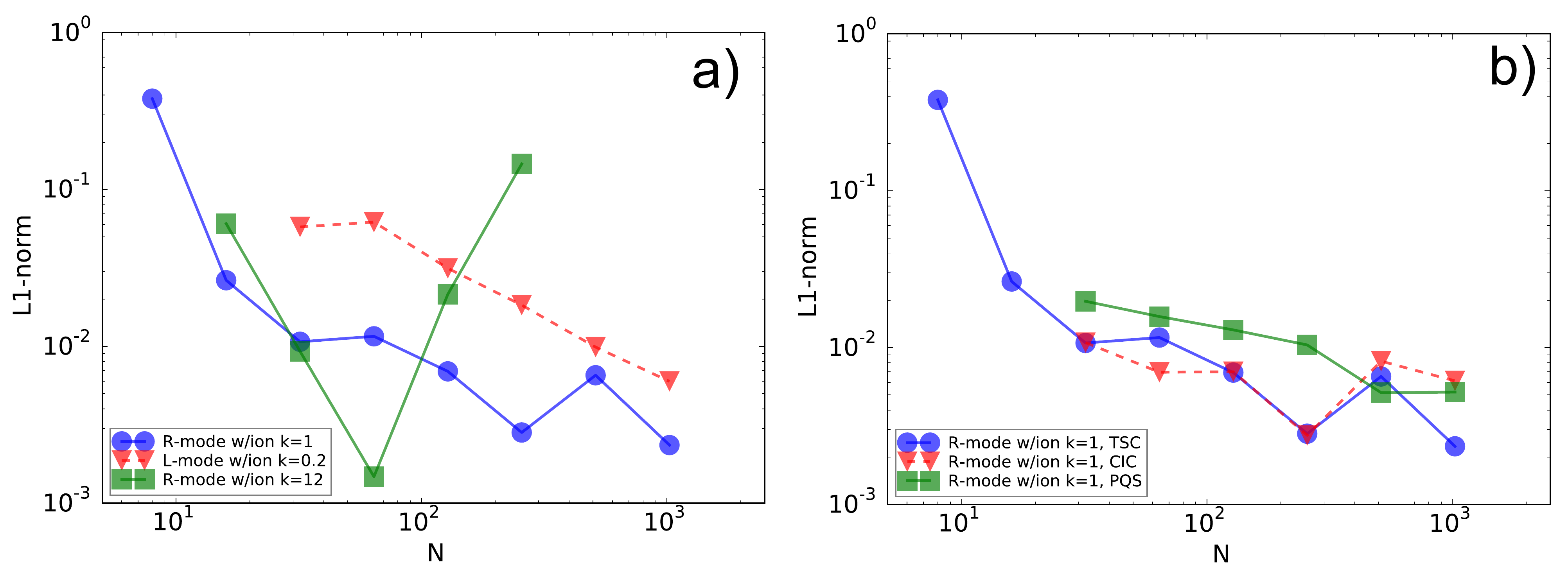}
	\caption{Convergence on N (inversely proportional to $\Delta x$) for: a) different waves: R-mode $kd_i=1$, L-mode $kd_i=0.2$,  R-mode $kd_i=12$. b) different interpolation methods for the same R-mode $kd_i=1$: linear (CIC), quadratic (TSC) and cubic (PQS) particle shape functions.
	\label{fig:convergence_dx_waves_interpolation}}
\end{figure}

The previously shown convergence tests were done only for R waves
with $kd_i=1$. We also test convergence on $\Delta x$ initializing other waves
with both low and high frequency.
In the first case, we chose an L mode with $kd_i=0.2$ (with a corresponding
wavelength of $\lambda=10\pi d_i$), where the Hall
effect is not important and the wave is almost dispersiveless, similar
to a pure MHD Alfv\'en wave.
The cell grid size ranges between $\Delta x/d_i=[0.03-0.98]$ and the
timestep is chosen similarly as above, fulfilling the CFL condition
at the same level for different cases.
For the second limit, we chose a high frequency whistler/R-mode wave
with $kd_i=12$ (with a corresponding wavelength of $\lambda=(\pi/6)\,d_i$),
in order to emphasize the finite phase speed of the whistler waves.
The cell grid size ranges between $\Delta x/d_i=[0.002-0.0005]$.
All the cases used full ion response with 1024 particle per cell.

The results are shown in Fig.~\ref{fig:convergence_dx_waves_interpolation}a),
where we also depicted the same R-mode $kd_i=1$ already analyzed
in Fig.~\ref{fig:n_convergence_whistler} for comparison purposes.
We infer that long-wavelength waves exhibit more numerical
error (e.g, red curve with $kd_i=0.2$) than a short-wavelength
(e.g, green curve with $kd_i=12$), for the same resolution
of the wave shape (same $N$).
The reason of this is that the relative ion/electron contribution to the wave shape
(associated to the current density of each specie) is more dominant
for the low frequency L mode than the high frequency R mode (dominated
by electron effects), with $kd_i=1$ in between.
Since the initial ion current distribution
in the initial magnetic profile is the main reason
for the numerical error in the wave shape (see previous Sec.~\ref{item2_ion}),
it is expected than any ion-dominated wave will be resolved
less accurately than an electron dominated wave
(see also Fig.~\ref{fig:n_convergence_whistler}).

There are other details in Fig.~\ref{fig:convergence_dx_waves_interpolation}a)
worth to explain.
The low frequency wave (red curve) exhibits a consistent convergence
(less numerical error) for increasing resolution in a more clear way
than the other high frequency waves. The reason for this behavior
is not clear at the moment, but is possibly due to the
initial larger grid cell sizes for the same $N$ compared to the other
short-wavelength waves.
We do not show the lower end of resolution for this L-wave,
because already for $N=32$ the resolution in $k$ space is so large
that also a wave in the R-branch is excited.
The two waves start to interfere and produced a distorted wave
shape,
departing
totally from the initial wave shape just after a few periods.
For comparison, the wave shape for the case of a R-mode with $kd_i=1$
is practically unchanged.
The L wave also shows a higher numerical error throughout
because it tends to excite parametrically ion sound waves,
modifying slightly the wave shape at short length-scales (seen as rippling).

On the other hand, for the high frequency R-mode in Fig.~\ref{fig:convergence_dx_waves_interpolation}a) (green curve),
an increasing resolution beyond $N=64$ leads to an increase in the
error, as a result of a second wave being generated, leading
to a beating and a complete
deformation of the wave shape. In this case, the wave amplitude actually increases.

\subsubsection{Interpolation effects on the convergence on $\Delta x$}

We also analyzed the effects of the interpolation scheme between
particle and fields (during the particle deposition and pusher),
in order to check for their correct implementation and effects
of a broader particle shape function in the accuracy of the algorithm.
For this sake, we repeat the aforementioned test of an R-mode for $kd_i=1$,
1024 particles per cell (and full ion response) for three
different shape functions: CIC (linear), TSC (quadratic) and PQS (cubic).
By default, all the previous tests used TSC.
The results are shown in Fig.~\ref{fig:convergence_dx_waves_interpolation}b).
The difference is in general small, in particular between the CIC
and TSC shape functions, providing evidence of the correct implementation
of those different schemes.
The higher order PQS shape function displays a slightly larger error
than the lower order shape functions.
This might be, perhaps, because it involves a broader particle
shape function covering more than a single cell,
causing a smoothing and modifying the wave shape, in particular
in regions of strong gradients.

\subsubsection{Simulated frequency and dispersion relation}

Finally, we also carried out a parametric study to obtain the
frequency $\omega_0$ selected by the system due to the launched wave
with a given wave number $k_0$. For this sake, we analyzed both L and R
wave modes, as well as the limit of a R-mode without ion response
(pure electron whistler). After four (theoretical) time periods,
a Fourier transform is applied to the time series $x-t$ and then
the resulting power in the $\omega-k$ dispersion relation is plotted
for the specific $k$ used in the initialization.
In such a way, a very
clear peak in the spectral power is obtained at the expected frequency.
This frequency also matches very well with the slope obtained in the
evolution of the wave profiles in the $x-t$ space,
which allows to derive a phase speed and from there the frequency.

\begin{figure}[!ht]
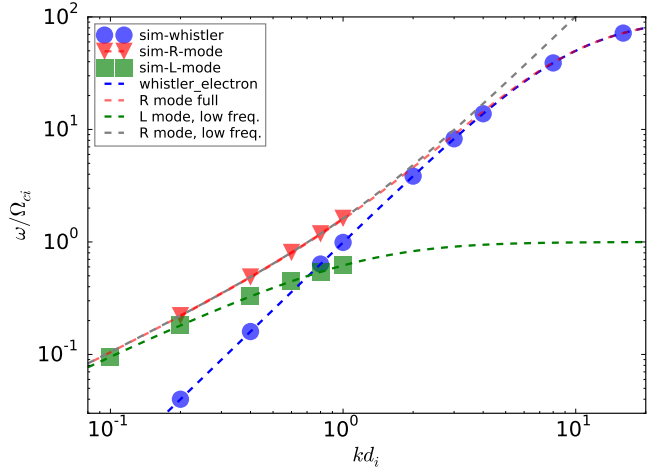
\centering
	\includegraphics[width=0.7\linewidth]{{{dispersion_omega_k}}}
	\caption{Simulated dispersion relation for whistler without ion effects, R and L modes.
	The dashed lines show the analytical expressions for the three waves, in addition to the R mode approximation
	for low frequencies (limit of $m_e\to 0$).
	\label{fig:dispersion_relation}}
\end{figure}

The results of this parametric study are shown in Fig.~\ref{fig:dispersion_relation},
together with the theoretical dispersion curves.
The agreement is very good for the 3 different curves,
although it is necessary to keep in mind
that the associated numerical errors in $d\omega$ are up to $1/4$ of the
theoretical frequency due to the running period.
The L mode waves for $kd_i\gtrsim 0.5$ are actually damped, and the
wave amplitude decreases with time, but a frequency can be clearly identified.
For wavenumbers larger than $kd_i\gtrsim 1.0$
the damping is so strong that it is difficult to assign a frequency.
Note also that the high frequency R (whistler) mode follows very close
the asymptote towards $\omega\to\Omega_{ce}$, implying a decreasing
phase speed with the associated numerical stabilization not
present in other hybrid codes without electron inertia.

\subsection{Parallel electromagnetic modes \label{sec:normal_modes}}

In order to reproduce the theoretical curves of the parallel electromagnetic modes
given by the dispersion relation Eq.~\eqref{eq:lr_mode},
without imposing any initial wave as in Sec.~\ref{sec:numerical_normal_modes},
we initialize a quasi-1D
(mostly along one direction, averaging over 4 cells in the transverse direction)
thermal plasma  with an external static magnetic field along the resolved $\hat{x}$-direction.
We do not apply any perturbation other than the (enhanced) thermal noise caused by the PiC shot noise
due to the reduced number of macro-particles.
This and similar plasma waves test problems (see also Sec.~\ref{sec:bernstein_modes}) are very useful for benchmarking of hybrid or PiC codes~\cite{Kilian2017}.
The physical and numerical parameters for the simulation are summarized in Tables~\ref{tab:physical_parameters_tests} and~\ref{tab:numerical_parameters_tests}, respectively.
Note that $V_A/c=\Omega_{ci}/\omega_{pi}$ and $\sqrt{\beta_i}=\sqrt{2}v_{th,i}/V_A$, where $v_{th,i}=\sqrt{k_BT_i/m_i}$ is the ion thermal speed, $T_i$ is the ion temperature and
$\beta_i$ is the ratio of ion thermal pressure to magnetic pressure.
The field output is written every 8 timesteps, resulting in a maximum captured output frequency of $\omega_{max}/\Omega_{ci}=\pi/(\Delta t\cdot8\cdot\Omega_{ci})= 78.5$,
a minimum resolved frequency of $\omega_{min}/\Omega_{ci}=2\pi/(T\Omega_{ci})= 0.0104$, and a maximum spatial resolution in the parallel wave number of $k_{x,min}d_i=0.06$.
The timestep $\Delta t$ and grid resolution $\Delta x$ are chosen to satisfy a CFL number for the electron Alfv\'en speed of $V_{Ae}\Delta t/\Delta x=0.5$. At the end of the simulation, the total energy is conserved within $1\cdot10^{-4}$ of its initial value.

\begin{figure}[!ht]
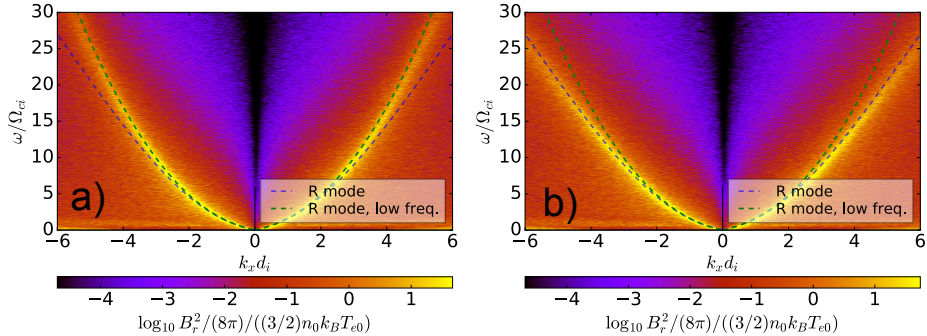
\centering
\includegraphics[width=0.99\linewidth]{{{fig3_dispersion}}}
	\caption{Right-handed (R) power spectra of perpendicular magnetic field for $B_y-iB_z$. a) inertia-less case $m_e=0$ and b) inertial case $m_e\neq0$. Dashed lines are the cold plasma approximations discussed in the text. \label{fig:r_modes_full}}
\end{figure}

The dispersion curves for the L and R modes can be obtained by a Fourier transform in space and time of the following complex combination of the components of the magnetic field.
\begin{equation}\label{eq:lr_decomposition}
	\sqrt{2}B_{R/L} = B_y\pm iB_z
\end{equation}
This choice corresponds to a change of basis vectors from Cartesian coordinates to vectors representing left- and right-handed circularly polarized waves (see Eq.~9.116 in Ref.~\cite{Baumjohann1997}, Sec.~1.4 in Ref.~\cite{Stix1992}, Sec.~5.2.1 in Ref.~\cite{Gary1993} or  Appendix~B of  Ref.~\cite{Terasawa1986}).
Thus, by plotting the spectral power in the magnetic field $B_R$ in the $\omega-k$ space,  we obtained the simulation dispersion curves for the cases of zero (Fig.~\ref{fig:r_modes_full}(a)) and finite electron mass (Fig.~\ref{fig:r_modes_full}(b)).
Our code reproduces correctly the cold full R-mode dispersion relation for both finite electron mass  (Eq.~\eqref{eq:lr_mode}) and massless electrons (Eq.~\eqref{eq:lr_lowfreq}). Note that the massless electron dispersion curve is accurate for low frequencies but it displays significant deviations from the cold full R-mode dispersion relation for $\omega\gg \Omega_{ci}$.
Note that the electron inertia effects are not only important near the electron cyclotron frequency $\Omega_{ce}=100\,\Omega_{ci}$, but also for much lower frequencies: the finite electron mass has significant effects in the dispersion curves for frequencies as low as  $\Omega_{ce}/5$, which demonstrates the importance to keep their effects even in this (relatively low) frequency regime.
Overall, this comparison shows that our code can handle accurately the cases of both zero and finite electron mass.

Note that in Fig.~\eqref{fig:r_modes_full} we did not show the higher frequency range close to the electron resonance $\Omega_{ce}$ because the available power in that region corresponding to higher $k$ decays exponentially (plot not shown here).
This observation agrees qualitatively with the known theoretical predictions of the electric field spectrum corresponding to the thermal fluctuations of the fully-kinetic plasma model~\cite{Langdon1970,Melzani2013}.

\begin{figure}[!ht]
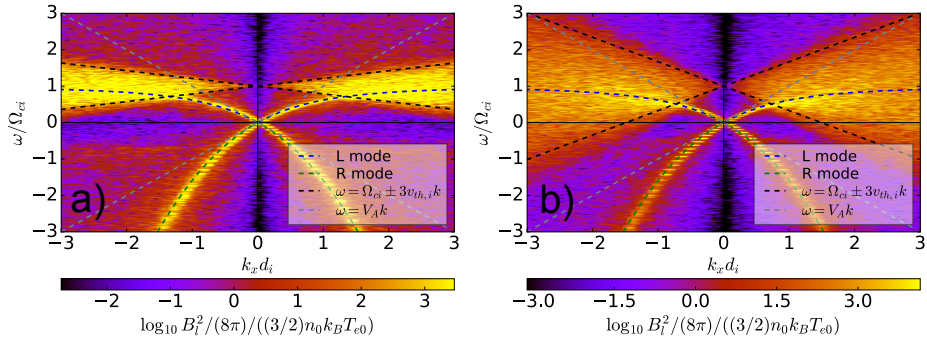
\centering
	\includegraphics[width=0.99\linewidth]{{{fig4_dispersion}}}
	\caption{Left-handed (L) power spectra of perpendicular magnetic field $B_L=B_y-iB_z$ for a)  $\beta_i=0.01$ (standard parameters) and b) $\beta_i=0.1$. Dashes lines are the cold plasma approximations discussed in the text. The curves with $\omega<0$ correspond to right-handed (R) polarized modes.  \label{fig:l_modes_zoom}}
\end{figure}

Since the L mode tends asymptotically to the ion cyclotron frequency $\Omega_{ci}$ for large $k$, it is convenient to study its behavior in the low frequency range. The results for the inertial case (there are practically no differences with the inertia-less case in this frequency range), are shown in Fig.~\ref{fig:l_modes_zoom}a). The agreement between the theoretical and simulation dispersion curves for the L mode (for $\omega>0$) is quite good as long as it does not enter into the triangular regions approximately delimited by
\begin{align}\label{eq:resonance_cones}
	\omega=\Omega_{ci} \pm 3\,v_{th,i}\,k .
\end{align}
The frequency $\omega$ and wave number $k$ satisfying this resonance condition represent thermal fluctuations caused by the ions. The factor ``$3$'' comes from the particles that are in the tail of the ion distribution function, three standard deviations from the mean (zero) velocity,
in agreement with our initial Maxwellian velocity distribution function.
The L modes gets heavily damped when it enters in this region, in agreement with the hot Vlasov dispersion relation (not shown here) and previous studies~\cite{Araneda2011}.

Note that Fig.~\ref{fig:l_modes_zoom}a) shows for $\omega<0$  the right-handed (R) polarized waves, since the polarization of the waves depends on the sign of $\omega$ (see Sec.~5.2.1 in Ref.~\cite{Gary1993}).
For very low frequencies $\omega\ll \Omega_{ci}$, both L and R modes follow the (MHD) Alfv\'en wave dispersion relation $\omega=V_A\,k$. The deviations observed for high frequencies (for $kd_i\gtrsim 0.5$) are due to the Hall effect, i.e., when the electron and ion dynamics need to be considered separately.

Fig.~\ref{fig:l_modes_zoom}b) shows the same power spectra in $B_R$  from another simulation run enhancing thermal effects by setting $\beta_i=0.1$, but otherwise exactly identical parameters
to the case shown in Fig.~\ref{fig:l_modes_zoom}a).
It shows the broadening of the triangular region due to the thermal fluctuations of ions in gyro-resonance in agreement with Eq.~\eqref{eq:resonance_cones}, because their thermal speed is larger. Note that the distribution of spectral power in Fig.~\ref{fig:l_modes_zoom}b) is more spread  inside of this triangular-like region compared to the previous lower-$\beta_i$ case Fig.~\ref{fig:l_modes_zoom}a).

\subsection{Perpendicular propagation: ion Bernstein and X-mode \label{sec:bernstein_modes}}

Bernstein waves are electrostatic plasma modes propagating nearly perpendicular to a magnetic field, coupled to gyrating particles~\cite{Bernstein1958}. They only exist in a plasma with finite temperature and have the remarkable property of being almost undamped and usually with a negative group velocity. Their dispersion curves are ordered in several bands at harmonics of the electron or ion cyclotron frequencies, with a general expression given by, e.g.,
Eq.~7.65 of Ref.~\cite{Boyd2003} or
Eq.~11.85 of Ref.~\cite{Stix1992}.
Because of the predicted high power at the harmonics of $\Omega_{ci}$, ion Bernstein modes are ideal to test cyclotron resonance effects of perpendicularly propagating waves in our code. Since we are interested in the frequency range $\Omega_{ci}<\omega\lesssim\Omega_{ce}$, the corresponding waves are called pure ion Bernstein modes, with a contribution coming mostly from ion kinetic effects. The electrons contribute only with an additional term. The relevant dispersion relations is then (see., e.g., Eq.~69 in Ref.~\cite{Rasmussen1994}):
\begin{align}\label{eq:bernstein}
	  & 1 - \frac{\omega_{pe}^2}{\Omega_{ce}^2} -  2\omega_{pi}^2\frac{e^{-\lambda_{i}}}{\lambda_{i}}\sum_{n=1}^{\infty}\frac{n^2}{\omega^2 - n^2\Omega_{ci}^2}I_n(\lambda_{i})=0.
\end{align}
where $I_n(\lambda_{i})$ is the modified Bessel function (of the first kind) of order $n$ and argument:
\begin{align}
	\lambda_{i} = k_{\perp}^2\rho_{i}^2,
\end{align}
where $\rho_{i}^2=v_{th,i}^2/\Omega_{ci}^2=k_BT_{i}/(m_{i}\Omega_{ci}^2)$ is the thermal ion  Larmor radius. Note that these waves are mostly independent of the electron temperature $T_e$. In the fluid limit $\lambda_{i}\to 0$, the upper bound in frequency tends to the lower hybrid oscillations $\omega=\Omega_{LH}$.

In order to excite these waves, we initialize a quasi-1D thermal plasma along the $\hat{x}$ direction
and an initial external static magnetic field along the $\hat{z}$ direction.
No perturbations are applied other than the PiC shot noise.
This is similar to the setup in Sec.~\ref{sec:normal_modes}, but changing the direction of the magnetic field.
The physical and numerical simulation parameters are summarized in Tables~\ref{tab:physical_parameters_tests} and~\ref{tab:numerical_parameters_tests}, respectively.
These parameters are slightly different to the ones used for the test problem in Sec.~\ref{sec:normal_modes}:  both plasma-$\beta_i$ and mass ratio have been increased to discern more easily the thermal properties of the Bernstein modes depending on $\rho_i$, as well as to isolate the pure ion effects by choosing a larger separation of scales between electrons and ions. Note that the numerical parameters for this problem, combined with the output frequency (every 242 timesteps), result in a maximum captured frequency of $\omega_{max}/\Omega_{ci}=\pi/(\Delta t\cdot242\cdot\Omega_{ci})= 31$,  a minimum resolved frequency of $\omega_{min}/\Omega_{ci}=2\pi/(T\Omega_{ci})= 0.0104$, and a maximum resolution in the parallel wave number of $k_{x,min}d_i=0.1$. The timestep $\Delta t$ and grid resolution $\Delta x$ have been chosen to give a CFL number for the electron Alfv\'en speed of $V_{Ae}\Delta t/\Delta x=0.25$.

\begin{figure}[!ht]
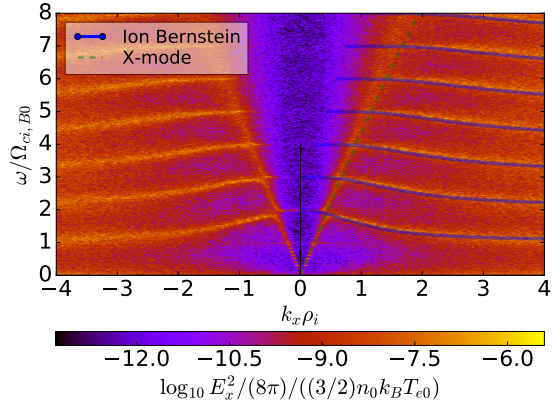
\centering

	\includegraphics[width=0.6\linewidth]{{{fig5_bernstein}}}
	\caption{Power spectrum of the perpendicular electric field $E_x$. Theoretical curves for the ion Bernstein modes  Eq.~\eqref{eq:bernstein} and X-mode are shown in the right-half ($k>0$) of this plot. \label{fig:bernstein}}
\end{figure}

Fig.~\ref{fig:bernstein} shows the spectral power of the parallel electric field $E_x$.
We can see a good agreement for all the harmonic of the fundamental mode $n=1$ of the pure ion Bernstein waves. Note that all the modes start close to $(n+1)\Omega_{ci}$ for small $k_{\perp}=k_x$ and end up at $n\Omega_{ci}$ for large $k_{\perp}$, with a negative group speed asymptotically decreasing towards large $k$.
We verified that smaller mass ratios produce additional gyro-resonances in thin bands at harmonics $n\,\Omega_{ci}$, on top of the pure ion Bernstein modes.\cite{Rasmussen1994}

In Fig.~\ref{fig:bernstein}, we also show the theoretical cold dispersion relation for the X-mode (see, e.g., Eq. 6.59 in Ref.~\cite{Boyd2003}). This is a cold plasma mode with perpendicular propagation and several branches at differences frequencies. In our case of interest, with parameters  suitable for the hybrid model, the maximum frequency of the X-mode is  the lower-hybrid frequency. The small disagreement seen in Fig.~\ref{fig:bernstein} between the simulation results and the theoretical (cold) dispersion curve can be explained due to the thermal effects neglected in the latter.

\subsection{Ion beam R instability}\label{sec:ionbeam}

A two (ion) component plasma is unstable to an instability driven by the relative drift speed between a main core (``c'') ion population and a more tenuous beam (``b'') component, along the  direction of a background magnetic field. For a relatively cold ion beam, the instability with lowest threshold produces unstable magnetosonic waves with right-handed polarization (see, e.g., Ref.~\cite{Winske1984}, Sec.~5.3.1 in Ref.~\cite{Matsumoto1993} or Sec.~8.2.1 in Ref.~\cite{Gary1993}). There are two different types of this instability: resonant and non-resonant.  In the first one, the ions produce resonant waves with phase speeds in the tail of the ion thermal distribution. The second one is fluid-like, driven by the bulk anisotropy of the ion distribution, and therefore much stronger than the resonant instability, because  most of the particles contribute to the interaction.

\begin{table}[!ht]
		\begin{tabular}{ccccccc}
			Case                  & $n_b/n_e$ & $U_{b-c}/V_A$ & $\beta_c=\beta_b$ & $T_c/T_b$ & $T_b/T_e$ & ppc-beam \\ \hline
			Resonant  R-beam & 0.02 & 10            & 1.0               & 1         & 1         & 512      \\
			Non resonant R-beam & 0.1 & 10            & 1.0               & 1         & 1         & 512
		\end{tabular}
	\caption{Main physical and numerical parameters for ion beam R instability. $c/b$ stands for beam/core related quantities. $n$ is density, $U_{b-c}$ is the relative beam-core drift speed.\label{tab:parameters_ionbeam}}
\end{table}

In order to trigger this instability, we consider a quasi-1D thermal plasma
with an external static magnetic field along the $\hat{x}$ direction.
We consider two populations of ions: the main core and a tenuous beam. The specific parameters of those two populations are summarized in Table~\ref{tab:parameters_ionbeam} for the two cases to be considered. The remaining parameters referred only to the ``core''-ions (and electrons) are specified in Tables~\ref{tab:physical_parameters_tests} and ~\ref{tab:numerical_parameters_tests}. The CFL number for the electron Alfv\'en speed is $V_{Ae}\Delta t/\Delta x=0.4$. Note that we chose only half of the number of particles per cell for the beam compared to the core, changing suitably its macrofactor (ratio of physical to numerical particles) to compensate for its lower density in order to
get better statistics.
We perform the simulation in the reference frame of the center of mass  of the ions, satisfying:
\begin{equation}\label{eq:center_of_mass}
	n_b\vec{U}_b + n_c\vec{U}_c = 0,
\end{equation}
where $n_{(b/c)}$ and $\vec{U}_{(b/c)}$ are the densities and bulk drift speeds of the beam/core ion populations in this reference frame, respectively. Eq.~\eqref{eq:center_of_mass}  leads to the explicit initial drift speeds of each specie in terms of their relative drift $\vec{U}_{b-c} = \vec{U}_b - \vec{U}_c$:
\begin{align}\label{eq:center_mass}
	\vec{U}_b & = \phantom{a} \frac{1}{1+n_b/n_c}\vec{U}_{b-c},  \\
	\vec{U}_c & = -\frac{n_b/n_c}{1+n_b/n_c}\vec{U}_{b-c}.
\end{align}
Note that in this reference frame the total ion current also vanishes.

\begin{figure}[!ht]
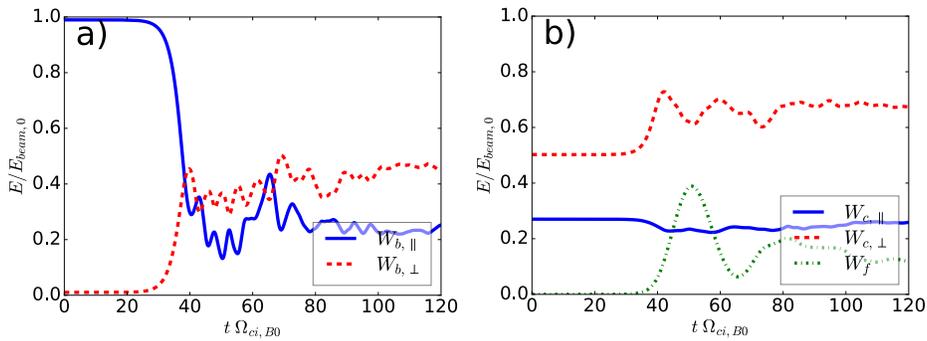
\centering
	\includegraphics[width=0.99\linewidth]{{{fig6_ionbeam}}}
	\caption{Time histories of the total particle energies for both ion beam and core. Resonant case. a) Beam particle parallel energy $W_{b,\parallel}$ and perpendicular energy $W_{b,\perp}$. b) Core particle parallel  $W_{c,\parallel}$, perpendicular energy $W_{c,\perp}$ and magnetic field energy $W_{f}$.\label{fig:ionbeam_energies_bothparticles}}
\end{figure}

Fig.~\ref{fig:ionbeam_energies_bothparticles} shows the energies of the  beam and core components for the resonant case.
The green curve in Fig.~\ref{fig:ionbeam_energies_bothparticles}(b)  shows the increase of the magnetic field energy due to the development of the instability.
We can see in Fig.~\ref{fig:ionbeam_energies_bothparticles}a) that the parallel beam drift speed diminishes very quickly during the linear growth phase of the instability,
and at the same time the perpendicular energy rises.
The latter corresponds mostly to a perpendicular beam heating due to the waves generated by the instability. Fig.~\ref{fig:ionbeam_energies_bothparticles}b) shows that the parallel core energy (consisting of field aligned thermal energy and the parallel bulk drift speed with respect to the center of mass) is mostly unaffected, while the core is heated in the perpendicular direction due to a mechanism similar to that of the beam. Our results agree with the ones shown in
Fig.~8 of Ref.~\cite{Winske1984}
and Fig.~4 of Ref.~\cite{Amano2014}.

\begin{figure}[!ht]
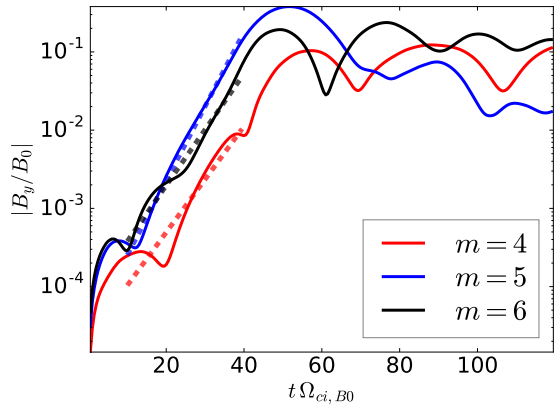

	\centering
	\includegraphics[width=0.6\linewidth]{{{fig7_ionbeam}}}
	\caption{Time history of the Fourier modes 4-6 of $B_y(k_x)$ for the resonant case. The dashed lines represent the exponential fitting used to calculate the growth rate $\gamma$ shown in Fig.~\ref{fig:ionbeam_gamma_k}(a). \label{fig:ionbeam_fftmodes}}
\end{figure}

In order to compare with the predictions of the linear theory~\cite{Winske1984,Matsumoto1993}, we calculate the three most unstable Fourier modes of the magnetic field component $B_y$. The results for the resonant case are shown in Fig.~\ref{fig:ionbeam_fftmodes}.
We determine the linear growth rate of those modes via an exponential fitting during the linear
phase of the instability,
indicated by dashed lines in Fig.~\ref{fig:ionbeam_fftmodes}.
Fig.~\ref{fig:ionbeam_gamma_k}(a) shows the growth rates  $\gamma$ of these and all the other unstable modes versus the wave number $k$. In Fig.~\ref{fig:ionbeam_gamma_k}(a) we also plot the theoretical growth rates  obtained by solving the hot Vlasov dispersion relation~\cite{Winske1984,Matsumoto1993}, illustrating very clearly the good agreement with our results.

\begin{figure}[!ht]
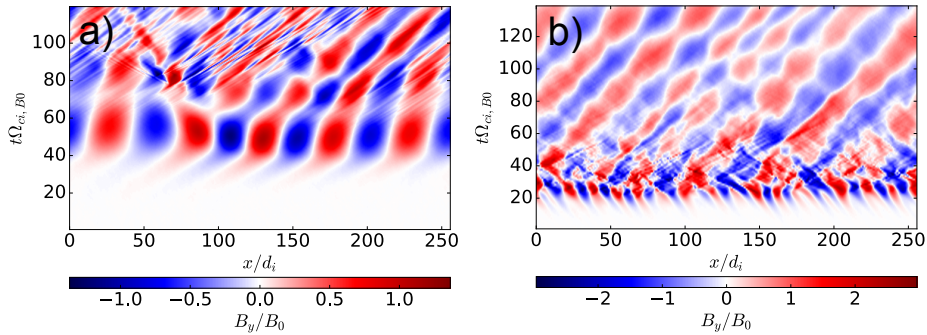

	\centering
 		\includegraphics[width=0.99\linewidth]{{{fig8_ionbeam}}}
	\caption{Stack time plot evolution of $B_y$. a) Resonant case. b) Non-resonant case \label{fig:ionbeam_stack}}
\end{figure}

\begin{figure}[!ht]
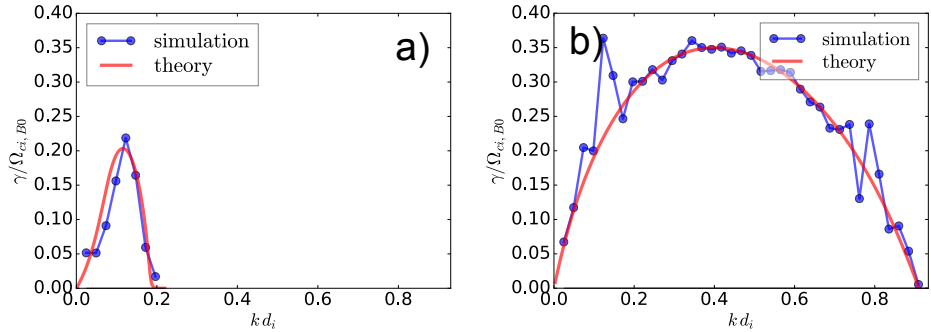

	\centering
 		\includegraphics[width=0.99\linewidth]{{{fig9_ionbeam}}}
	\caption{Growth rates $\gamma$ vs $k$ for the Fourier components of $B_y$. a) Resonant case. b) Non-resonant case. The theoretical curve was obtained by solving the hot plasma dispersion relation corresponding to these parameters. \label{fig:ionbeam_gamma_k}}
\end{figure}

The stack plot of $B_y$ in Fig.~\ref{fig:ionbeam_stack}a) confirms that, during the linear phase, there is a clear dominant mode with five full wavelengths in the simulation box. This plot also shows the preferentially forward propagating waves (to the right), as expected from the linear theory.

\begin{figure}[!ht]
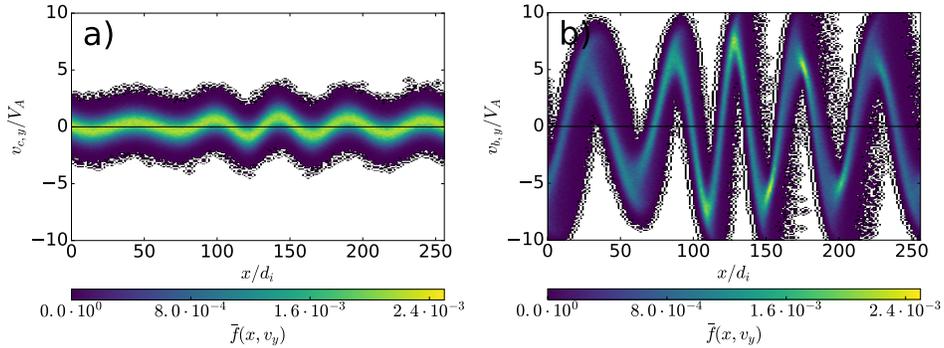
\centering
	\includegraphics[width=0.99\linewidth]{{{fig10_ionbeam}}}
	\caption{Ion core (a) and beam (b) phase space $x-v_y$, at $t\Omega_{ci}=40$. Resonant case. The normalized distribution function $\bar{f}$ is shown, satisfying $\iint f \,d\vec{x}\,d\vec{v}=1$. \label{fig:ionbeam_phasespace}}
\end{figure}

The dominant oscillation with five full wavelengths also affects both core and beam population. This can be seen in the corresponding phases spaces $x-v_y$ (perpendicular velocity) shown in Fig.~\ref{fig:ionbeam_phasespace}. Note the much stronger spreading of the beam population compared to the core. A significant number of beam particles reach perpendicular speeds close to $7V_A$, just  below the initial (parallel) drift speed ($10V_A$). This is qualitatively in agreement with previously published results for similar parameters
(see, e.g., Fig.~5 of \cite{Winske1984}, Fig.~3 of \cite{Amano2014}).

The results for the stronger, fluid-like, non-resonant case also match with the linear theory, as seen in Fig.~\ref{fig:ionbeam_gamma_k}(b). This plot shows the growth rate of the first 38 Fourier modes of $B_y$, up to the maximum unstable wave number predicted by the theory ($kd_i\approx 1$).
In Fig.~\ref{fig:ionbeam_stack}(b) we can also see the evolution of the spatial structure of $B_y$, showing much smaller wavelengths at maximum growth rates than in the resonant case. Note that the modes propagate to the left (backward waves) during the initial linear phase, also in agreement with the predictions of the linear theory. Only much later, after saturation, the dominant modes propagate forward (to the right). Overall, the amplitude of these modes is much larger than in the resonant-case, as expected from a stronger instability.

\subsection{Ion Landau damping}

One of the most characteristic kinetic plasma processes is the Landau damping~\cite{Landau1946}, where an electrostatic wave (from the ion-acoustic branch) is damped by transferring its energy to resonant particles with parallel speed $v_{\parallel}$ satisfying the resonance condition $\omega = v_{\parallel} k$.
Close to the ion frequencies $\Omega_{ci}$ and length scales $d_i$, this process can be observed
for long wavelengths $k\lambda_{De}\ll 1$ and for the range of phase speeds $v_{th,i} < v_{\phi} < v_{th,e}$, because this implies a weak damping for electrons hotter than the ions ($T_e \gg T_i$).
This is equivalent to $|\xi_i|\gtrsim 1$ for the (cold) ions and $|\xi_e| \ll 1$ for the (hot) electrons, where
\begin{align}\label{z_plasma_argument}
	\xi_{(i,e)}=\frac{\omega}{k\sqrt{2}v_{th,{(i/e)}}}=\frac{\omega_r+i\gamma}{k\sqrt{2}v_{th,{(i/e)}}}.
\end{align}
Here, $\omega_r$ is the real part of the frequency and $\gamma$ is the imaginary part (damping rate
if it is negative).
The hot dispersion relation that describes accurately the ion-acoustic waves subject to ion Landau damping, in the hybrid approximation, is given by~\cite{Kunz2014}:
\begin{align}\label{electrostatic_dispersion_hybrid}
	Z'(\xi_i) =-2\frac{T_i}{T_e}.
\end{align}
where $Z'$ is the derivative of the plasma zeta function~\cite{Fried1961}:
\begin{align}\label{z_plasma}
	Z(\xi_{i})=\frac{1}{\sqrt{\pi}}\int\limits_{-\infty}^{\infty}\frac{\exp(-x^2)}{x-\xi_{i}}dx,\qquad\mathfrak{Im}(\xi_{i})>0.
\end{align}
Note that Eq.~\eqref{electrostatic_dispersion_hybrid} involves only $\xi_i$ and not $\xi_e$, since kinetic effects are considered only for ions in the hybrid approximation.
In order to demonstrate Landau damping with our hybrid code, we setup a quasi-1D simulation along the $\hat{x}$ direction with no magnetic field, and a density perturbation of the form:
\begin{equation}
	n(x)= n_0(1 + \epsilon\cos(k_m x)),\qquad k_m =\frac{2\pi m}{L_x}.
\end{equation}
where $L_x$ is the simulation length box and $m$ the excited mode number. The parameters specific of this test problem are given in Table~\ref{tab:parameters_landaudamping} while the remaining physical and numerical parameters are in Table~\ref{tab:physical_parameters_tests} and Table~\ref{tab:numerical_parameters_tests}.
\begin{table}[!ht]
\centering
		\begin{tabular}{ccccccc}
			Case               & $\epsilon$ & $m$ \\ \hline
			ion Landau damping & 0.03           & 4   \\
		\end{tabular}
	\caption{Main numerical parameters for the ion Landau damping test problem.  We vary the ratio $T_i/T_e$ by changing $T_e$ and always keeping $T_i$ fixed.\label{tab:parameters_landaudamping}}
\end{table}

\begin{figure}[!ht]
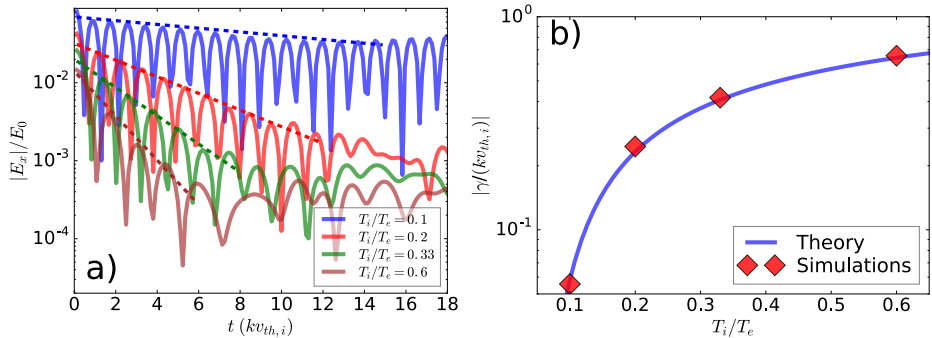
\centering
	\includegraphics[width=0.99\linewidth]{{{fig11_landau}}}
	\caption{Landau damping test problem. Time evolution (a) and comparison of the theoretical and simulation growth rates (b) of the Fourier mode $m=4$ for the electrostatic field $E_x$ and different temperature ratios $T_i/T_e$. In a), the dashed lines are the exponential fits from which the damping rates were calculated. \label{fig:ionlandau_damping}}
\end{figure}

The results of this test problem are summarized in Fig.~\ref{fig:ionlandau_damping}. The left panel shows the time evolution of the Fourier mode number $m=4$ of the electrostatic field $E_x$. We can see very clearly that by decreasing the temperature ratio $T_i/T_e$ the damping becomes weaker, and the initially imposed wave can propagate for longer time without losing (significant) power. This is in agreement with the theoretical expression Eq.~\eqref{electrostatic_dispersion_hybrid}.
Fig.~\ref{fig:ionlandau_damping}(a) also shows the damping rate calculated with an exponential fitting during the linear stage of the process, until the power of the initial wave reaches the noise floor. The growth rates are plotted against the temperature ratio $T_i/T_e$ in Fig.~\ref{fig:ionlandau_damping}(b) as red diamonds. The theoretical approximation for the damping rates, Eq.~\eqref{electrostatic_dispersion_hybrid}, is plotted as a blue curve. The agreement with the theory is very good, even considering that our code uses a full $f$-PiC approach for the ions, different from less noisy approaches such as hybrid-Vlasov codes~\cite{Valentini2007}, $\delta f$-hybrid PiC codes~\cite{Kunz2014}, or codes using the Vlasov-Hybrid simulation (VHS) method~\cite{Nunn1993}.

It is important to mention that simulations with a hybrid PiC code in the regime $T_e\gg T_i$   are prone to a numerical instability and/or numerical heating. As found by Refs.~\cite{Rambo1995,Rambo1997}, the regime of high electron temperature might excite an instability analogous to the well-known finite grid instability in fully-kinetic PiC codes~\cite{Birdsall1991}. In the latter case, the finite grid instability is due to the interaction of plasma oscillations (Langmuir waves) with the aliases of the spatial grid whenever the grid cell size is much larger (within a factor of 2-3) than the Debye length. The typical growth rates are on the order of the plasma frequency, with the electrons being heated until the electron temperature reaches a value such as the Debye length becomes on the order of the grid cell size. Similarly, in hybrid-PiC codes, the finite grid instability is due to ion-acoustic waves whenever the temperature ratio $T_e/T_i\gg 1$ (no sharp threshold), making difficult to choose parameters (in this regime) that makes a simulation stable within a reasonable computational cost. This instability heats ions until the ion thermal speed becomes comparable to the ion acoustic speed (and therefore, $T_e/T_i\sim 1$). Refs.~\cite{Rambo1995,Rambo1997} found that there will always be a numerical heating in the regime  $T_e/T_i> 1$. Smoothing and use of higher order particle shape functions helps to keep this effect efficiently controlled, and to a much smaller extent the increase of macro-particles per cell. This is the reason to not simulate scenarios with even higher electron temperature than the case $T_e/T_i=10$ shown in Fig.~\ref{fig:ionlandau_damping}.

\subsection{Firehose instability}\label{sec:firehose}

Temperature anisotropy can be a source of free energy for instabilities. In the hybrid regime,  for a sufficiently large plasma-$\beta_i$ and for $T_{i,\parallel}>T_{i,\perp}$, the corresponding fastest growing instability is called firehose (see, e.g., Sec.~3.4 of Ref.~\cite{Treumann2001a}). Although it is possible to describe this instability in (anisotropic) MHD or two-fluid models, its growth rate and maximum wave number for the case of oblique propagation (with respect to a background magnetic field) is correctly reproduced only by considering ion kinetics, in particular finite Larmor radius effects~\cite{Yoon1993}.
\begin{table}[!ht]
		\begin{tabular}{ccccccc}
			Case                     & $\beta_{i,\perp}=\beta_i$ & $\beta_{\parallel}$ \\ \hline
			ion firehose instability & $300/\pi$               & 3+$\beta_{i,\perp}$
		\end{tabular}
	\caption{Main physical parameters for the ion firehose test problem. \label{tab:parameters_firehose}}
\end{table}

We reproduce this instability with our code by initializing a quasi-1D simulation of a high beta plasma, with the external static magnetic field along the $\hat{x}-$direction. The specific parameters are given in Table~\ref{tab:parameters_firehose} while the remaining physical and numerical parameters are in Table~\ref{tab:physical_parameters_tests} and Table~\ref{tab:numerical_parameters_tests}.

\begin{figure}[!ht]
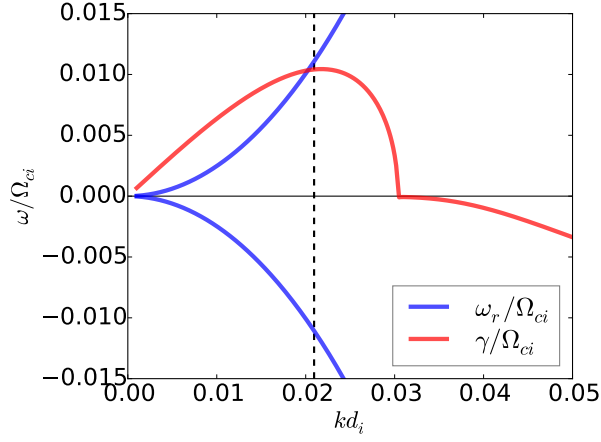
\centering
	\includegraphics[width=0.65\linewidth]{{{fig12_firehose}}}
	\caption{Frequency and growth rate of the hot plasma dispersion relation for the firehose parameters.  The black vertical dashed line represents the wave number $k$ associated to the mode $m=1$. \label{fig:firehose_dispersion}}
\end{figure}

\begin{figure}[!ht]
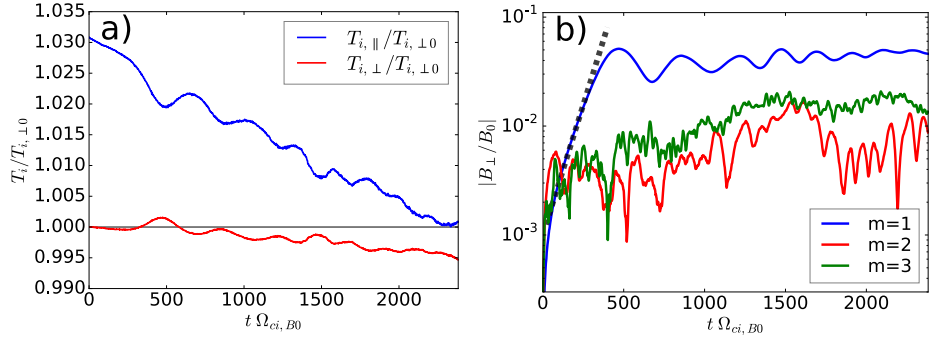
\centering
	\includegraphics[width=0.99\linewidth]{{{fig13_firehose}}}
	\caption{Time evolution of quantities related with the firehose instability test problem. a) Components of the globally averaged thermal ion speed: $v_{th,i,\alpha}^2=(1/N)\sum_j^N \left(v_{\alpha,j}-V_{\alpha}\right)^2$, with the component $\alpha=x,y,z$ for the $j$th ion of a total of $N$ and $V_{\alpha}$ is the $\alpha-$component of the bulk (average) ion speed. b) Fourier modes of the transverse magnetic field $B_{\perp}(k_x)=\sqrt{[B_y(k_x)]^2 + [B_z(k_x)]^2}$. \label{fig:firehose_timeevolution}}
\end{figure}

Our results are summarized in Fig.~\ref{fig:firehose_timeevolution}. The left panel shows that the initially higher values of the parallel ion temperature (proportional to $v_{th,i,x}^2$) decrease in time reaching similar values to those of the perpendicular ion temperature  (associated with $v_{th,i,y}^2$ or $v_{th,i,z}^2$). In this way, the source of free energy of the instability is exhausted. This instability produces transverse magnetic fluctuations during its linear stage, whose first three Fourier modes are shown in Fig.~\ref{fig:firehose_timeevolution}(b). According to the linear theory~\cite{Kunz2014}, the most unstable mode should be $m=1$ with growth rate $\gamma=0.011\Omega_{ci}$, and similarly for the real frequency $\omega$ (both positive and negative). The solution of the dispersion relation for both $\gamma$ and $\omega$ is plotted in Fig.~\ref{fig:firehose_dispersion}, where we also indicate the mode number 1 in our configuration. This clearly shows that our setup allows to develop only a single mode, because all the harmonics that our simulation box allows are  damped. Note that the real frequency curve has both positive and negative branches, indicating that the instability generates a superposition of both forward and backward propagating waves.
Fig.~\ref{fig:firehose_timeevolution}(b) shows this fact and also that the theoretical $\gamma$ fits very well with the simulated growth rate of this instability.

\subsection{Oblique (2D) firehose instability}\label{sec:firehose2d}

Ion temperature anisotropies with $T_{i,\parallel}>T_{i,\perp}$ can also drive (firehose) instabilities with waves propagating obliquely to a background magnetic field (AF, oblique propagating Alfv\'en firehose). In a 2D geometry and under suitable parameters, both (firehose) instabilities with waves propagating obliquely and parallel to the magnetic field (see Sec.~\ref{sec:firehose}, denoted onwards as  WF: parallel propagating Whistler-firehose) can coexist with similar growth rates. The oblique firehose instability was found for the first time by Ref.~\cite{Hellinger2000} and simulated together with the parallel firehose by Ref.~\cite{Hellinger2001}. Different from the parallel propagating firehose known since much earlier, the oblique firehose instability generates waves with zero frequency that can isotropize more efficiently the ion temperature anisotropy.

In this section, we present a 2D test problem showing the simultaneous existence of this kind of instabilities and their identification based on their different properties, following Ref.~\cite{Hellinger2001}. By comparing with the linear and non-linear theory, this will allow us to prove that our code can handle the spectral transfer of power between waves with different propagation angles and the cyclotron resonances effects on the ion distribution function. We reproduce this instability with our code by initializing a 2D simulation with the external static magnetic field along the $\hat{x}-$direction. The specific parameters are given in Table~\ref{tab:parameters_firehose_2d} while the remaining physical and numerical parameters are in Table~\ref{tab:physical_parameters_tests} and Table~\ref{tab:numerical_parameters_tests}. These parameters give a comparable maximum growth rate of $\gamma/\Omega_{ci}\sim 0.056$ for the parallel firehose instability (WF) and $\gamma/\Omega_{ci}\sim 0.059$ for the oblique firehose instability (AF)~\cite{Hellinger2000}.

\begin{table}[!ht]
		\begin{tabular}{ccccccc}
			Case                     & $T_{i,\perp}/T_{i,\parallel}$ \\ \hline
			2D ion firehose instability(ies) & $0.4$
		\end{tabular}
	\caption{Main physical parameters for the ion firehose test problem. \label{tab:parameters_firehose_2d}}
\end{table}

\begin{figure}[!ht]
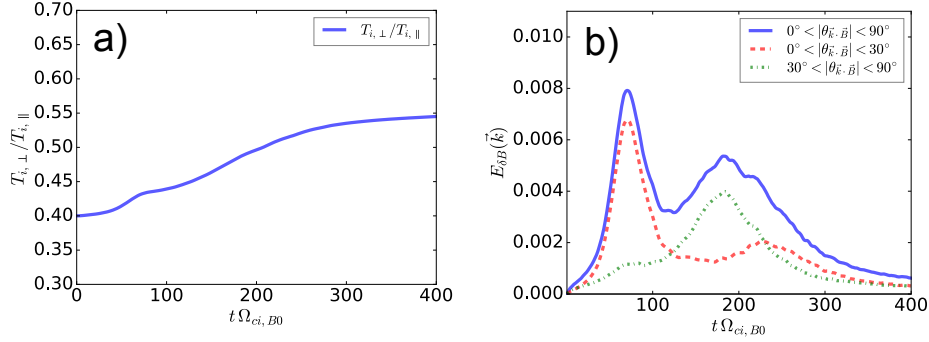
\centering
	\includegraphics[width=0.99\linewidth]{{{fig14_firehose2d}}}
	\caption{Time evolution of the firehose 2D test problem. a) Ion anisotropy $T_{i,\perp}/T_{i,\parallel}$. b) Spectral power of the total magnetic field fluctuating energy $E_{\delta B}$ and its components: near parallel propagation ($0^{\circ}<\theta_{\vec{k}\cdot\vec{B}}<30^{\circ}$) and oblique propagation ($60^{\circ}<\theta_{\vec{k}\cdot\vec{B}}<90^{\circ}$)). \label{fig:firehose2d_evolution}}
\end{figure}

Our results are summarized in Fig.~\ref{fig:firehose2d_evolution} with the time evolution of two quantities characterizing the system. Fig.~\ref{fig:firehose2d_evolution}a) shows the drop in the initial imposed ion temperature anisotropy (globally averaged), exhausting the source of free energy of the firehose instability(ies). Most of the anisotropy is reduced in the time interval between $100<t\Omega_{ci}<200$, to later remain practically constant. This origin of this behavior is clearly explained in Fig.~\ref{fig:firehose2d_evolution}b) with the time evolution of the spectral power in the total magnetic field fluctuating energy $E_{\delta B}(\vec{k})$ and its components aligned and oblique to the background magnetic field. This splitting requires to define the angle  in the Fourier space $\theta_{\vec{k}\cdot \vec{B}}={\rm acos}(k_x/\sqrt{k_x^2+k_y^2})$ between the wave number vector  $\vec{k}$ and the background magnetic field in $x$. The power in a given interval of angles is obtained then just by integrating all the power in $\delta B^2$ which falls into that range. In this way, Fig.~\ref{fig:firehose2d_evolution}b) shows that most of the magnetic fluctuations are initially propagating nearly parallel to the magnetic field. These fluctuations, reaching a peak near $t\Omega_{ci}\sim 100$, are due to the parallel propagating firehose instability (WF). Later,   between $100\lesssim t\Omega_{ci}\lesssim 200$, the power in the parallel fluctuations drops considerably while the nearly oblique waves start to develop, contributing to most of the total power near $t\Omega_{ci}\sim 200$. Later, until the end of the simulation at $t\Omega_{ci}\sim 400$, the power in both parallel and oblique propagating waves decreases considerably. All these features match very closely with the results published in Ref.~\cite{Hellinger2001} and are explained to a greater extent there.

\begin{figure}[!ht]
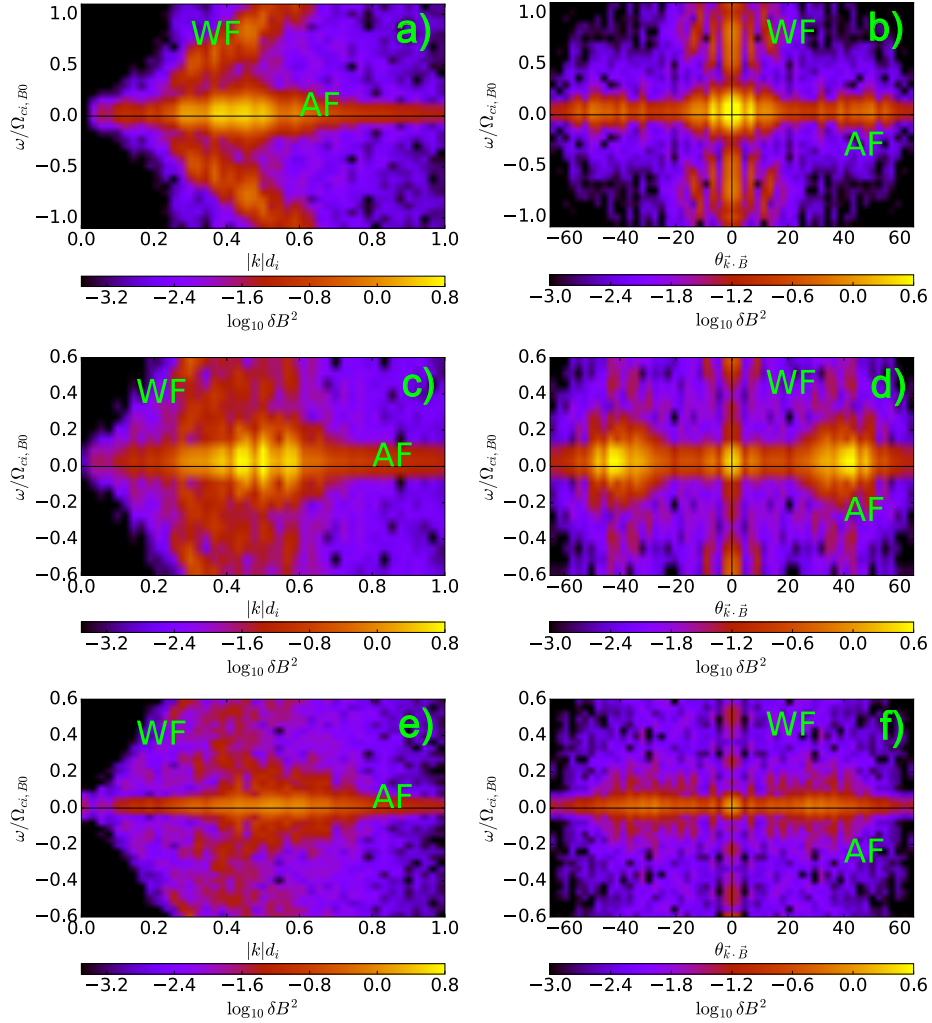
\centering
	\includegraphics[width=0.99\linewidth]{{{fig15_firehose2d}}}
	\caption{Dispersion relations of the power in the magnetic field fluctuating energy of the firehose 2D test problem. Left: $\omega-|k|$. Right: $\omega-\theta_{\vec{k}\cdot\vec{B}}$. First row: initial, between $0<t\Omega_{ci}<100$. Second row: middle, between $100<t\Omega_{ci}<200$. Third row, later, $200<t\Omega_{ci}<400$. WF stands for the parallel propagating Whistler-firehose, while AF stands for oblique propagating Alfv\'en firehose. \label{fig:firehose2d_dispersions}}
\end{figure}

Fig.~\ref{fig:firehose2d_dispersions} shows the distribution of spectral power in the $\omega-|k|$ (left column) and  $\omega-\theta_{\vec{k}\cdot\vec{B}}$ (right column) planes for the different phases of spectral power transfer between parallel and oblique waves identified before. For early times $0<t\Omega_{ci}<100$, Fig.~\ref{fig:firehose2d_dispersions}(b) shows the power in the parallel propagating firehose (WF) at $\theta_{\vec{k}\cdot\vec{B}}\sim 0^{\circ}$ and with a frequency near $\omega/\Omega_{ci}\sim \pm 0.6$. This branch follows the whistler R Alfv\'en mode with finite phase speed as seen in  Fig.~\ref{fig:firehose2d_dispersions}(a) near $|k|d_i\sim 0.6$. On the other hand, the  oblique waves generated by the oblique Alfv\'en firehose instability (AF) are seen at $\theta_{\vec{k}\cdot\vec{B}}\sim 45^{\circ}$ in  Fig.~\ref{fig:firehose2d_dispersions}(b), at nearly zero frequency, and with a similar wave number as for the parallel propagating WF. There is also power at $\theta_{\vec{k}\cdot\vec{B}}\sim 0^{\circ}$ and with $|k|d_i\sim 0.6$. All these features agree very well with the solutions of the Vlasov dispersion relation for these parameters (see Figs.~2 and 4 in Ref.~\cite{Hellinger2000}).

Later, between  $100<t\Omega_{ci}<200$,  Fig.~\ref{fig:firehose2d_dispersions}(d) shows that power is transferred to the oblique waves of the AF instability as evidenced by an enhancement of power there and a decrease in the WF branch after its saturation. This is consistent with the behavior seen in  Fig.~\ref{fig:firehose2d_evolution}b). Finally, between $200\lesssim t\Omega_{ci}\lesssim 400$, after the saturation of the AF instability, Fig.~\ref{fig:firehose2d_dispersions}f) shows very little power in both AF and WF branches (in particular the latter one), as a result of the exhaustion of the ion anisotropy which drives these instabilities.

\begin{figure}[!ht]
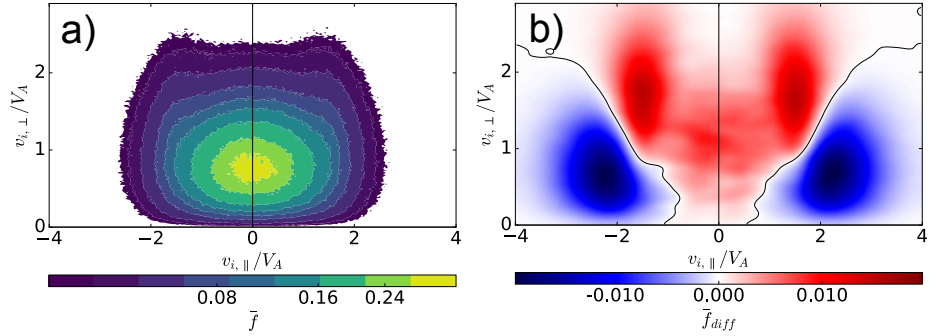
\centering
	\includegraphics[width=0.99\linewidth]{{{fig16_firehose2d}}}
	\caption{Ion VDF of  the firehose 2D test problem in the $v_{\perp}-v_{\parallel}$ plane. a) Final VDF at $t\Omega_{ci}=400$  b) Difference between the final and initial VDFs. The black line represents the contour level with no changes. \label{fig:firehose2d_vdfs}}
\end{figure}

Fig.~\ref{fig:firehose2d_vdfs} shows the effects on the ion velocity distribution function (VDF) of the different wave activity generated by the two firehose instabilities. In particular, Fig.~\ref{fig:firehose2d_vdfs}a) shows the final ion VDF at $t\Omega_{ci}=400$ with two peaks at $v_{\perp}/V_A\sim2$ and $|v_{\parallel}|/V_A\sim1$. The origin of these non-Maxwellian features can be traced back in Fig.~\ref{fig:firehose2d_vdfs}b), which shows the difference between the final and initial VDFs. There is an increase in the number of particles (red contours) precisely in the velocity space region mentioned before with the non-Maxwellian features between $v_{\perp}/V_A\sim2$ and $|v_{\parallel}|/V_A\sim1$.  These particles come from the region where there is a decrease (blue contours) near $v_{\perp}/ V_A\sim0.5-1$ and $|v_{\parallel}|/V_A\sim 2$. This is evidence that these particles were accelerated in the perpendicular direction via cyclotron resonance $\omega =k_{\parallel}v_{\parallel,res} \pm \Omega_{ci}$, a well known effect of the damped waves due to the AF instability. Indeed, as shown previously, this instability is characterized for waves with $\omega\sim 0\ll\Omega_{ci}$. Then, for $k_{\parallel}d_i=0.5-0.7$ (see Fig.~\ref{fig:firehose2d_dispersions}(d)), the resonant phase speed can be estimated to be in   $v_{\parallel,res} \sim 1.4-2V_A$, in agreement with the modified regions in the velocity space $v_{\parallel}$ in Fig.~\ref{fig:firehose2d_vdfs}b). Note that this cyclotron resonant effect does not happen in the first stage of the evolution of the system dominated by the WF instability, because their unstable waves do not satisfy the resonant condition.

\section{Conclusion}\label{sec:conclusion}

We presented a new hybrid code with kinetic ions modeled with the PiC method and inertial electrons modeled with the electron fluid equations without approximation. Our code was validated through a set of one numerical and six different physical test problems, covering different aspects of the  ion kinetic dynamics and electron inertia.

0) Excitation of parallel eigenmodes to check for convergence and stability of the underlying hybrid algorithm.

1) Parallel electromagnetic and 2) perpendicular electrostatic modes with the purpose of checking the excitation of normal plasma modes arising from thermal noise. We reproduce that the R mode and the ion Bernstein modes are mostly undamped, while the L mode gets damped when entering the dispersive regime for $k>d_i$ in the low frequency regime ($\omega<\Omega_{ci}$). The L mode and ion Bernstein modes follow the correct dispersion curves with practically pure ion contribution. The R mode has the correct contribution from both kinetic ions and inertial electrons, especially significant in the high frequency regime $\omega\gg\Omega_{ci}$. The latter is because it differs by a large amount from the curve followed by the model with kinetic ions and massless electrons, which is also contained in the code as a limiting case.

3) Ion beam R instability to check that the code reproduces correctly the evolution of instabilities having a relative drift speed between two ion populations as its source of free energy, and correctly handling multiple ion species in general.

4) Ion Landau damping to prove that the code can reproduce correctly the damping of electrostatic waves via wave-particle interactions.

5) Ion firehose instability to check the correct development of instabilities with ion temperature anisotropy as their source of free energy.

6) 2D oblique ion firehose instability to check the simultaneous presence of instabilities with waves propagating at different angles with respect to the background magnetic field in a 2D geometry. This also allows to verify the expected power transfer between these waves during the non-linear stage of these instabilities, in addition to verify perpendicular ion acceleration due to ion cyclotron resonance.

Altogether, these tests show that our simulation code can reproduce correctly the relevant ion kinetic physics in the regime $\omega\sim\Omega_{ci}$ and $k\sim d_i^{-1}$, by comparing with the predictions of linear (and non-linear) theory. Our hybrid code does no approximations for the electron inertial contributions appropriate for $\omega\sim\Omega_{ce}$ and $k~\sim d_e^{-1}$, other than the equation of state relating electron pressure and temperature. Therefore, the presented simulation approach is ideal to study dissipative or multi-scale phenomena between electron and ion scales, like collisionless shocks, magnetic reconnection and kinetic turbulence in laboratory, space, and astrophysical plasmas.

\section*{Acknowledgments}
	The authors acknowledge the developers of the ACRONYM code (Verein zur F\"orderung kinetischer Plasmasimulationen e.V.) which was used as part of our hybrid code. P.M. and J.B. acknowledge the financial support by the Max-Planck-Princeton Center for Plasma Physics (project MMCAERO8003). P.K. acknowledges support from the NRF and DST of South Africa through the following disclosure:\\
	This work is based upon research supported by the National Research Foundation and Department of Science and Technology. Any opinion, findings and conclusions or recommendations expressed in this material are those of the authors and therefore the NRF and DST do not accept any liability in regard thereto.\\
	All authors thank the referees for their very valuable and constructive
	comments which helped to improve the presentation of our results.


\end{document}